\documentclass[twocolumn]{aastex631}

\usepackage{amsmath}

\begin{document}

\title{Robustness of Neural Ratio and Posterior Estimators to Distributional Shifts for Population-Level Dark Matter Analysis in Strong Gravitational Lensing}

\author[0000-0003-4701-3469]{Andreas Filipp}\email{andreas.filipp@umontreal.ca}
\affiliation{Ciela Institute, Montreal Institute for Astrophysics and Machine Learning, Montreal QC, Canada}
\affiliation{Mila - Quebec AI Institute, Montreal QC, Canada}
\affiliation{Department of Physics, University of Montréal, Montréal QC, Canada}

\author[0000-0002-8669-5733]{Yashar Hezaveh}
\affiliation{Ciela Institute, Montreal Institute for Astrophysics and Machine Learning, Montreal QC, Canada}
\affiliation{Mila - Quebec AI Institute, Montreal QC, Canada}
\affiliation{Department of Physics, University of Montréal, Montréal QC, Canada}
\affiliation{Center for Computational Astrophysics, Flatiron Institute, NY, USA}
\affiliation{Perimeter Institute for Theoretical Physics, Waterloo, Canada}
\affiliation{Trottier Space Institute, McGill University, Montréal, Canada}

\author[0000-0003-3544-3939]{Laurence Perreault-Levasseur}
\affiliation{Ciela Institute, Montreal Institute for Astrophysics and Machine Learning, Montreal QC, Canada}
\affiliation{Mila - Quebec AI Institute, Montreal QC, Canada}
\affiliation{Department of Physics, University of Montréal, Montréal QC, Canada}
\affiliation{Center for Computational Astrophysics, Flatiron Institute, NY, USA}
\affiliation{Perimeter Institute for Theoretical Physics, Waterloo, Canada}
\affiliation{Trottier Space Institute, McGill University, Montréal, Canada}

\begin{abstract}
We investigate the robustness of neural ratio estimators (NREs) and sequential neural posterior estimators (SNPEs) to distributional shifts in the context of measuring the abundance of dark  matter subhalos using strong gravitational lensing data.
While these data-driven inference frameworks can be accurate on test data from the same distribution as the training sets, in real applications, it is expected that simulated training data and true observational data will differ in their distributions.
We explore the behavior of a trained NRE and trained SNPEs to estimate the population-level parameters of dark matter subhalos from a large sample of images of strongly lensed galaxies with test data presenting distributional shifts within and beyond the bounds of the training distribution in the nuisance parameters (e.g., the background source morphology).
While our results show that NREs and SNPEs perform well when tested perfectly in distribution, they exhibit significant biases that often lead to not recovering the ground truth in the $3\sigma$ interval when confronted with slight deviations from the examples seen in the training distribution. 
This indicates the necessity for caution when applying NREs and SNPEs to real astrophysical data, where high-dimensional underlying distributions are not perfectly known.
\end{abstract}


\section{Introduction} 
One of the most striking open problems in modern astrophysics is that the nature of $\sim 80\%$ of the matter content of the universe remains unknown \citep[e.g.,][]{Davis_1985, Blumenthal_1985, White_1987, DM_review_2005}. This form of matter, called dark matter, is believed to consist of new, invisible particles that do not interact with regular matter electromagnetically. Shedding light on the nature of the dark matter particle is one of the main goals of modern cosmology and particle astrophysics \citep[e.g.,][]{Drlica_2019_LSSTDM}.

It has been well understood that different dark matter particle properties can result in different spatial distributions of dark matter structures on various scales \citep[e.g.,][]{Kuhlen_2012}. Thus, by measuring the spatial distribution of dark matter it is possible to discriminate between different dark matter models. 

The most commonly accepted dark matter model, cold dark matter (CDM) has been able to explain the observations of the large scale structure (e.g., the cosmic microwave background, baryon acoustic oscillations, weak lensing) with great precision and accuracy \citep[e.g.,][]{WMAP_2013, Planck_2020}.
However, on subgalactic scales a number of discrepancies between the predictions of CDM and the observations of the dwarf satellites of the Milky Way have given rise to the possibility of alternative dark matter models \citep[e.g.,][]{Kravtsov_2012_review}. 

Strong gravitational lensing, the formation of multiple images of distant light sources due to the deflection of their light rays by the gravity of intervening structures, is a powerful probe of the subgalactic distribution of matter in the lensing galaxies and along the line of sight to the background sources due to its purely gravitational nature.
Since different spatial distributions for the projected matter density can result in different distortions in the images, the analysis of lensed images allows for the inference of these projected densities, including the abundance and distribution of subhalos. These can then be related to population-level parameters such as the halo mass function on these subgalactic scales, which are directly linked to predictions of dark matter models.

However, measuring the effect of small halos on lensed images is a challenging, nonlinear inverse problem. The signal is weak and suffers from multiple degeneracies with other nuisance parameters, such as the morphology of the background source. 
Furthermore, the properties of a population of dark matter subhalos correspond to a high-dimensional space (e.g., the positions and masses of a large number of subhalos), making the inference of the abundance and distribution of subhalos a difficult problem for traditional, explicit likelihood modeling methods. 
Past works have introduced a number of approximations in an attempt to make the problem tractable, for example by assuming Gaussian priors, linearizing the lensing model, limiting the analysis to only modeling the effect of the most massive subhalos, or performing a power-spectrum analysis \citep[e.g.,][]{Vegetti_2010, Yashar_2016, Hezaveh_2016_powerspec, Cyr_Racine_2016, Birrer_2017_summary_stats, Brennan_2019, Despali_2020_lensing_prop_subhalos}. 
Despite these simplifications, these methods are still generally computationally costly, which limits the possibility of extensive testing them for potential biases.  

Neural network inference frameworks, such as neural ratio estimators (NREs) and sequential neural posterior estimators (SNPEs), have recently emerged as a promising solution to these problems since they can be trained to approximate the intractable likelihood or the posterior of parameter distributions directly from high-dimensional input data \citep[e.g.,][]{Cranmer_2015_NRE, Baldi_2016_NRE, Papamakarios_2016_NDE, He_2016_NPE, brehmer2018constraining, brehmer2019mining, Brehmer_NRE_particle_2020}. 
In principle, both analysis frameworks can marginalize over large numbers of nuisance parameters and return an optimal, unbiased likelihood or posterior for the parameters of interest. 
For more details on NREs and SNPEs and their motivation in likelihood free inference see Appendix~\ref{app:nlfim}.

Within the context of subhalo studies with strong gravitational lensing, the networks infer parameters describing the population-level distribution of subhalos (e.g., the subhalo mass function) for a collection of strong lensing systems, while marginalizing over nuisance parameters, including the source galaxy morphologies, the macro-lens parameters, and the parameters of an a priori unknown number of individual subhalos. 
These methods predict the subhalo population parameters for each lens individually; afterward, the predictions of the individual lenses are combined to get a population-wide posterior.
Recent work has demonstrated the promise of such approaches to circumvent approximations of the intractable likelihood in strong gravitational lensing \citep[e.g.,][]{Brehmer_Sidd_2019_NRE, Coogan_2022, Sidd_2022_NRE, Zhang_2022, Eve_NRE_2023, Karchev_2023, Wagner_C2023, Wagner_C2024, Zhang_2024}. 

Despite their potential, these methods face challenges when applied to real observational data.
While these methods have been shown to work well when tested on data coming from the same distribution as the training data, their performance on out-of-distribution (OOD) data is not guaranteed. 
These subhalo inference frameworks rely on producing large volumes of labeled, simulated data for training, requiring a detailed match between the distributions of real and simulated data for unbiased inference. However, even using the most sophisticated simulation pipelines, it is to be expected that there will always remain some level of mismatch between synthetic and real data distributions. Given the weakness of the expected signal of interest, this leaves SNPEs and NREs vulnerable to biased inference. 
The degrading of network performances on OOD data is a challenge beyond the field of astrophysics \citep[e.g., ][]{Adversarial_attacks_2013, Nguyen_NN_OOD_misspec, Cannon_modelmisspec_SBI}.

In this work, we investigate the performance of NREs and SNPEs for the inference of the subhalo mass function in the presence of realistic distributional shifts between the training and test datasets. We produce several simulated datasets with minor variations in background source morphologies, the distribution of lens macro model parameters, subhalo profiles, and observational noise statistics and show that when the training and test datasets are distributionally shifted, the performance of the NRE and SNPE can be dramatically affected. Note that these experiments are not meant to explore an exhaustive list of possible distributional shifts, but rather to establish the vulnerability of this class of methods to such shifts in strong lensing analysis. 
Further, we do not aim to qualify or rate the induced biases and qualitatively explore which modifications worsen the bias effect, but rather we aim to show quantitatively that OOD shifts bias the inferred posteriors in an unpredictable way and lead to untrustworthy inference results.

In Section~\ref{sec:method} we describe the methods and the inference frameworks. In Section~\ref{sec:data_generation} we describe the data generation process for the inference frameworks. In Section~\ref{sec:tests}, we present the details of the distributional shifts considered. In Section~\ref{sec:results} we present the performance of the inference frameworks on OOD tests and present our conclusions in Section~\ref{sec:Conclusion}. In what follows, we adopt a flat $\Lambda$CDM cosmology with parameters from the \citet{Planck_2020}.

\section{Methods}
\label{sec:method}
\subsection{Bayesian Inference of Dark Matter Models}
In strong lensing systems, a background source described by the parameter vector $S$ is distorted by a mass distribution in the foreground. 
For example, the source could be represented as a Sérsic light profile with $S$ containing the Sérsic parameters, or as a pixelated image with $S$ containing the pixel values.

The foreground mass density is considered to be the sum of two components: a smooth profile for the main deflector and small-scale, local fluctuations resulting from low-mass dark matter subhalos. The smooth component is typically represented by an analytic profile, such as the singular isothermal ellipsoid (SIE) profile, with $L$ containing the parameters of this profile. Individual dark matter subhalos, $m_i$, are also often represented with simple analytic profiles, like the Navarro-Frank-White (NFW) profile \citep[][]{NFW_profile}, or a truncated NFW profile (tNFW) with a finite density expansion in comparison to the NFW profile. However, since there are a large number of these subhalos, $H=\{m_1,m_2,...\}$ contains these parameters for the entire subhalo population (e.g., the positions, masses, and truncation radii of a large number of subhalos). Also note that, since the number of subhalos is unknown, this is a transdimensional problem, and $H$ could have different lengths for different models.  

The parameters of individual subhalos, $m_i$ (i.e., their masses $m_{h,i}$ and their position $\psi_i$ in the image plane), are drawn from a population-level distribution with parameters $\vartheta$ (for us the normalization and the slope of the dark matter subhalo mass function), which are the parameters of interest for our studies. We are interested in the posterior distribution of these subhalo population parameters, marginalized over all nuisance parameters, $S$, $L$, $H$, etc. For simplicity, we collectively denote all the nuisance parameters as $\theta = \{ S,L,H \}$ and our data as $D$.  

By using many independent observations $\{D_i\}$, we can get a population-wide posterior estimation of our parameters of interest $\vartheta$ from the observations $\{D_i\}$.

To obtain the posterior distribution $p(\vartheta | \{D_i\})$ of the parameters of interest $\vartheta$ given a set of independent observed images $\{D_i\}$, we apply Bayes' theorem:
\begin{equation}
\begin{split}
    p(\vartheta|\{D_i\}) &= \frac{p(\vartheta) \prod_i p(D_i|\vartheta)}{\int \text{d}\vartheta' p(\vartheta') \prod_i p(D_i|\vartheta')} \\
    &= p(\vartheta) \left[ \int \text{d}\vartheta'p(\vartheta') \prod_i \frac{p(D_i|\vartheta')}{p(D_i|\vartheta)} \right]^{-1} 
    \label{eq:posterior}
\end{split}
\end{equation}
where $p(\vartheta)$ is the prior on the population-level distribution parameters.

The models of the observations $\{D_i\}$ explicitly depend on the intermediate macro model parameters $\theta$, which themselves depend on the population-level parameters $\vartheta$. 
Thus, the likelihood $p(D_i|\vartheta)$ is the marginalization of the likelihood $p(D_i,\theta|\vartheta)$ over all intermediate model parameters $\theta$:
\begin{equation}
    p(D_i|\vartheta) = \int \text{d}\theta p(D_i,\theta|\vartheta) 
    \label{eq:goal_likelihood}
\end{equation}
The likelihood $p(D_i|\vartheta)$ and the posterior $p(\vartheta|\{D_i\})$ are intractable due to the high-dimensional nature of the likelihood $p(D_i,\theta|\vartheta)$, which can be expressed as:
\begin{equation}
    \begin{split}
        p(D,&\theta|\vartheta) = p_{\rm lens}(L) \\ 
        &\quad \times \text{Pois}(n_{h}|\bar{n}_{h}(\vartheta)) \prod_{i=1}^{n_h}\left[p_{\rm mass}(m_{h,i}|\vartheta) p(\psi_i)\right] \\
        &\quad \times p_{\rm obs}(D|\theta) 
    \end{split}
    \label{eq:simplify_likelihood}
\end{equation}
with $p_{\rm lens}$ the prior of the lens parameter distribution $L$, and $\bar{n}_h(\vartheta)$ the expected number of subhalos in the strong lens as a function of the parameter of interest $\vartheta$, and $n_h$ is the realized number in a specific simulation. 
As later defined (in Section~\ref{sec:train_nre} and \ref{sec:train_snpe}), in this paper the parameters of interest are for the NRE $\vartheta = (f_{\rm sub}, \beta)$, and define the dark matter population parameters of the subhalo abundance $f_{\rm sub}$ and the mass slope $\beta$ of the mass function, and for the SNPE $\vartheta = \Sigma_{\rm sub}$ defines the dark matter mass function normalization drawn from a mean $\Sigma_{\rm sub, pop}$ with standard deviation $\Sigma_{\rm sub, pop, \sigma}$.
$m_{h,i}$ and $\psi_i$ are the realized masses and positions of the subhalos in the simulation (i.e., the individual subhalo parameters $m_i = \{m_{h,i}, \psi_i \}$), with $p(\psi_i)$ the prior on the subhalo positions. $p_{\rm mass}$ is the normalized subhalo mass function, and $p_{\rm obs}$ is the probability of observing the specific image $D$ based on the lens parameter configuration $\theta$, taking Poisson fluctuations and the point spread function (PSF) into account \citep[see e.g.,][]{Brehmer_Sidd_2019_NRE}. Note that the contribution of the subhalos to the likelihood assumes statistical independence between the different subhalos.

Both the NRE and the SNPE effectively marginalize over the nuisance parameters by learning the intractable likelihood, enabling the posterior calculation from Equation~\ref{eq:posterior}.

\subsection{Neural Ratio Estimator}
Rather than directly learning the likelihood from the intractable integral in Equation~\ref{eq:goal_likelihood} or the posterior from Equation~\ref{eq:posterior}, the training objective of NRE is the likelihood-to-evidence ratio $r(D, \theta|\vartheta)$:
\begin{equation}
    r(D,\theta|\vartheta) = \frac{p(D,\theta|\vartheta)}{p_{\rm ref}(D, \theta)}
    \label{eq:ratio}
\end{equation}
where $p_{\rm ref}(D, \theta)$ is the evidence, defined as the reference likelihood of an observation $D$ occurring under any possible dark matter parameterization $\vartheta$. 
\begin{equation}
    p_{\rm ref}(D,\theta) = \int \text{d}\vartheta' p(\vartheta')p(D, \theta|\vartheta')
\end{equation}
Here, $p(D, \theta|\vartheta')$ is the likelihood of an observation $D$ given a specific set of dark matter parameters $\vartheta'$, and $p(\vartheta')$ is the prior distribution for the dark matter parameters used in training data generation.

Predicting the likelihood-to-evidence ratio $r(D,\theta|\vartheta)$, rather than the intractable likelihood or the posterior, allows for more efficient training of the neural network. 
The ratio simplifies the likelihood in Equation~\ref{eq:simplify_likelihood} to:
\begin{equation}
    \begin{split}
        r(D,\theta|&\vartheta) = \\
        &\text{Pois}(n_{h}|\bar{n}_{h}(\vartheta)) \prod_{i=1}^{n_h}\left[p_{\rm mass}(m_{h,i}|\vartheta) p(\psi_i) \right] 
    \end{split}
\end{equation}
because all other terms cancel out when taking the likelihood-to-evidence ratio, as they do not depend on the parameters of interest $\vartheta$.
The ratio can be computed analytically and used as training objective, when the mass function of the dark matter subhalos is analytic, and the probability for spatial distributions is straightforward to compute, as discussed in \citet{brehmer2018constraining, brehmer2019mining, Brehmer_Sidd_2019_NRE}.
By training the network on images $D$ generated with different sets of parameters $(\theta, \vartheta)$ and the corresponding likelihood-to-evidence ratio $r(D,\theta|\vartheta)$, the NRE learns to marginalize over the nuisance parameters $\theta$.

The trained NRE predicts the marginal likelihood-to-evidence ratio $r(D|\vartheta)$, which can be used to calculate the posterior from Equation~\ref{eq:posterior} by replacing the fraction of likelihoods with the fraction of marginal likelihood-to-evidence ratios:
\begin{equation}
    \frac{p(D_i|\vartheta')}{p(D_i|\vartheta)} = \frac{\frac{p(D_i|\vartheta')}{p_{\rm ref}(D_i)}}{\frac{p(D_i|\vartheta)}{p_{\rm ref}(D_i)}} = \frac{r(D_i|\vartheta')}{r(D_i|\vartheta)}
\end{equation}
This substitution reduces the posterior calculation to
\begin{equation}
    p(\vartheta|\{D_i\}) = p(\vartheta) \left[\int \text{d}\vartheta' p(\vartheta') \prod_i \frac{r(D_i|\vartheta')}{r(D_i|\vartheta)} \right]^{-1}
\end{equation}
The parameter space that needs to be integrated over is defined by the parameters of interest $\vartheta$. For our NRE, this is a two-dimensional space (e.g., the normalization and slope of the subhalo mass function, discussed in Section~\ref{sec:train_nre}), making the integral in the posterior calculation low-dimensional and tractable.

In the limit of unlimited training data from the same distribution as the test data and perfect learning, the NRE is capable of predicting the marginal likelihood-to-evidence ratio perfectly. However, in practice, these conditions are unattainable. 
To ensure the reliability and accuracy of the NRE at the time of inference, the trained NRE is calibrated. 
The calibration ensures that the NRE’s outputs — the predicted likelihood-to-evidence ratios — correctly match the true frequencies of parameters in practice.
This calibration process addresses potential biases, overconfidence or underconfidence, and discrepancies in the uncalibrated outputs produced by the NRE. The calibration procedure follows the method described in \citet{brehmer2019mining, Brehmer_Sidd_2019_NRE}.

\subsection{Sequential Neural Posterior Estimator}
In comparison to the NRE, the SNPE predicts the intractable posterior (Equation~\ref{eq:posterior}) directly. 
For the SNPE, a network is initially trained on a broad prior and predicts Gaussian posteriors (i.e. a mean with uncertainty) for the parameters of interest as well as for a number of nuisance parameters -- in our case the Einstein radius, the mass slope of the elliptical power law (EPL), the lens position, the ellipticity of the lens, the external shear, and the position of the source -- to make the learning process easier. 
The whole ensemble of parameters with predictions from the SNPE are denoted by $\varphi$.
The inferred posteriors on these parameters define priors to fine-tune the neural network by further training with the updated training priors \citep[e.g., ][]{Wagner_C2023, Wagner_C2024}.

For population inference, a set of sequential networks must be trained for each observation $D_i$ independently.
Every target image requires the training of an independent set of SNPE networks.

The distribution of the training data for the SNPE effectively serve as a prior during inference. A distributional shift between the training data and the test data can result in biased inference. However, this bias can be corrected for in the hierarchical inference procedure as described in \citet{Wagner_C2021}.
For this, the SNPE outputs -- the estimated posterior $q_\phi(\varphi|D_k, \vartheta_{\rm prior})$ -- can be incorporated into a hierarchical model to calculate the posterior of the \textit{observed} parameters of interest $\vartheta_{\rm obs}$ given a number $N_{\rm lens,obs}$ of observed lenses $\{D\}$ and the assumed \textit{prior} on the parameters of interest $\vartheta_{\rm prior}$. The prior on the parameters of interest gets updated with every sequential step of the SNPE.
For the hierarchical modeling, we follow \citet{Wagner_C2023, Wagner_C2024}:
\begin{equation}
\begin{split}
    p(\vartheta_{\rm obs}|\{D_{i}\}) = p(\vartheta_{\rm obs}) \times \prod_{k=1}^{N_{\rm lens,obs}} \frac{p(D_k|\vartheta_{\rm prior})}{p(\{D\})} \times \\
    \prod_{k=1}^{N_{\rm lens,obs}} \int \text{d}\varphi \frac{p(\varphi|\vartheta_{\rm obs})}{p(\varphi|\vartheta_{\rm prior})} q_\phi(\varphi|D_k, \vartheta_{\rm prior})
    \label{eq:hierarchical_inf}
\end{split}
\end{equation}
The third term represents an importance sampling integral. Equation~\ref{eq:hierarchical_inf} re-weights the posteriors to account for the differences between the prior distribution $\vartheta_{\rm prior}$ and the underlying distribution in the observation $\vartheta_{\rm obs}$. 

There is no general analytical solution to Equation~\ref{eq:hierarchical_inf}, but the three distributions in the integral, $p(\varphi|\vartheta_{\rm obs})$, $p(\varphi|\vartheta_{\rm prior})$, and $q_\phi(\varphi|D_k, \vartheta_{\rm prior})$, are by construction Gaussian. This special configuration provides an analytical solution to the integral and allows the use of a sampling algorithm to draw from $p(\vartheta_{\rm obs}|\{D_{\rm i}\})$. We use a Markov Chain Monte Carlo sampler from the \texttt{emcee} package \citep{emcee}. More details on the hierarchical inference and importance sampling can be found in \citet{Wagner_C2021}.

\section{Data Generation} \label{sec:data_generation}
We test the frameworks of NRE and SNPE for subhalo inference using two publicly available implementations by \citet{Brehmer_Sidd_2019_NRE}\footnote{\url{https://github.com/smsharma/mining-for-substructure-lens}} and \citet{Wagner_C2024}\footnote{\url{https://github.com/swagnercarena/paltax/tree/main}}, respectively. 
These two models use simulations with different parameterizations of the dark matter subhalo population and use different code bases for the simulation of the strong lensing systems. 
These methods are existing works, trying to probe specifially the signal of DM substructure, therefore we use their suggested parameterizaitons \citep[see][]{Brehmer_Sidd_2019_NRE, Wagner_C2024}.
Hence, the different implementations of the networks do not allow for an easy adaptation to match the parameterizations. The tests in this work are based on the reproduction of existing work -- in the case of the NRE with a modification of the used lensing simulator to fit our needs better -- with extensions to test different distributional shifts. 
Below we describe the baseline simulations (i.e. the training sets) for each model, followed by a description of the OOD test data in Section~\ref{sec:tests}.  

In both frameworks the individual dark matter halos are sampled from a subhalo mass function that accounts only for the subhalos in the main deflector. In this work we do not account for halos in the line of sight or other effects on the subhalo mass function beyond a power-law.

    \subsection{Training Simulations for the NRE} \label{sec:train_nre}
    To simulate the lenses in the NRE framework, we use \href{https://github.com/Ciela-Institute/caustics}{\texttt{caustics}}\footnote{\url{https://github.com/Ciela-Institute/caustics}} \citep[][]{Caustics_2024}. 
    
    We model the light of the background source with an ensemble of 5 to 50 Sérsic profiles \citep[][]{Sersic_1963}, to model background sources with different complexities. 
    
    The main deflector is an SIE profile, with the normalized surface mass density $\kappa$
    \begin{equation}
        \kappa(x, y) = \frac{3-\gamma_{\rm epl}}{2} \left(\frac{\theta_{\rm E}}{\sqrt{q x^2 + y^2/q}} \right)^{\gamma_{\rm SIE}-1}
        \label{eq:EPL_kappa}
    \end{equation}
    with $\theta_{\rm E}$ the Einstein radius, $\gamma_{\rm SIE} = 2.0$ the mass slope of the SIE, and $q$ the axis ratio of the mass profile. $(x,y)$ define a Cartesian coordinate system aligned with the major and minor axis of the mass profile \citep[][]{Tessore_EPL}.
    For the case of $\gamma_{\rm SIE} \neq 2.0$ the SIE profile becomes a more general EPL profile, which is used in one of the test cases.

    The Einstein radius $\theta_{\rm E}$ of the EPL depends on the mass within the Einstein radius $M(\theta_{\rm E})$
    \begin{equation}
        \theta_{\rm E} = \sqrt{\frac{4GM(\theta_{\rm E})}{c^2} \frac{D_{\rm ls}}{D_{\rm l} D_{\rm s}}}
    \end{equation}
    and defines the typical separation scale between multiple images.  

    The individual dark matter halos are modeled with NFW profiles \citep[][]{NFW_profile}.
    The NFW profile can easily be modified to a tNFW profile by adding a truncation radius to the NFW profile, at which the mass distribution of the profile gets cutoff. This is used in on of our test cases.
    The mass of each subhalo is sampled from a mass function.
    The standard model of cosmology, the $\Lambda$CDM model, predicts  a scale-invariant power spectrum of primordial fluctuations. From this, subhalos in galaxies are expected to follow to first order a power-law mass distribution. 
    The power-law mass function is described by the following equation:
    \begin{equation}
    \frac{\text{d}n}{\text{d log } m_{\rm halo}} = \alpha \cdot M_{\rm host} \cdot m_{\rm halo}^{\beta}
    \label{eq:dm_mf}
    \end{equation}
    with $\alpha$ the normalization constant of the profile, $M_{\rm host}$ the mass of the host galaxy, and $m_{\rm halo}$ the subhalo mass.
    To quantify and normalize the abundance of dark matter, we introduce the parameter $f_{\rm sub}$, which determines the ratio of mass in the dark matter subhalos to the mass in the main deflector defined by the SIE profile and defines the normalization $\alpha$:
    \begin{equation}
    \begin{split}
        f_{\rm sub} &= \alpha \int \text{d}m_{\rm halo} m_{\rm halo}^{\beta+1} \\
        &= \frac{\int \text{d}m_{\rm halo} m_{\rm halo} \frac{\text{d}n}{\text{d}m_{\rm halo}}}{M_{\rm host}}
    \end{split}
    \end{equation}
    This enables efficient sampling of dark matter subhalos across different dark matter mass functions. The parameters $(f_{\rm sub}, \beta)$ fully define the dark matter population. We aim to infer these two parameters with the NRE.
    The prediction from $\Lambda$CDM for this power-law approximation is $f_{\rm sub}\approx0.05$ and $\beta \approx -0.9$ \citep[e.g.,][]{Diemand_2007_ViaLactea, Kuhlen_2007_ViaLactea, Springel_2008, Hiroshima_2018}.

\begin{figure*}[ht]
        \centering
        \includegraphics[width=1.0\linewidth]{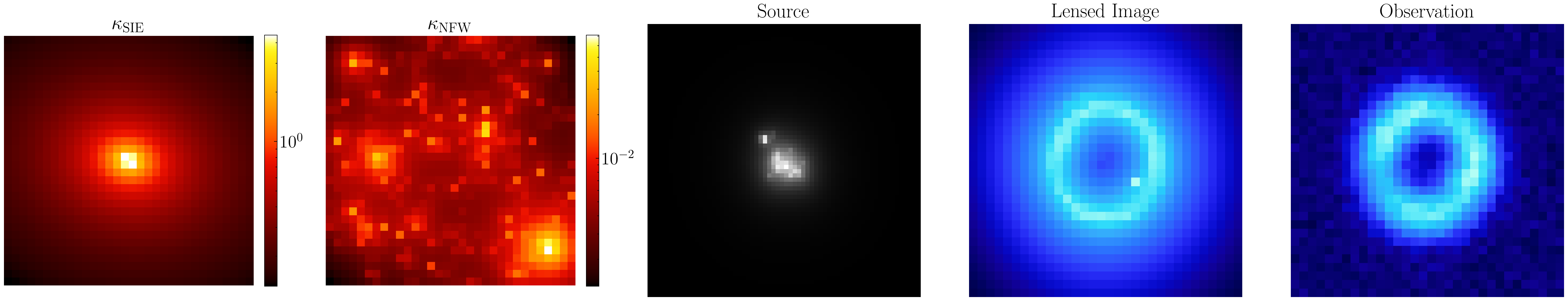}
        \caption{From left to right: a convergence map $\kappa_{\rm SIE}$ created with an SIE profile, the convergence map of a sample of dark matter halos $\kappa_{\rm NFW}$, a realization of the source light, the lensed image without PSF convolution and noise, and the lensed image with PSF convolution and noise, generated with the provided kappa maps and source light.}
        \label{fig:simulation_maps}
    \end{figure*}

    Figure~\ref{fig:simulation_maps} shows an example of all relevant components in the simulations. From left to right, the figure displays the convergence map $\kappa_{\rm SIE}$ generated by the main deflector -- an SIE profile --, the convergence map for a sample of dark matter halos $\kappa_{NFW}$,
    a realization of the source light, which consists of multiple Sérsic profiles, followed by the lensed image generated with the provided kappa maps and source light without PSF convolution, and finally the observation with PSF convolution and noise.

    The data specifications to train and test the NRE are based on the expected LSST data quality in the r-band after completion of the full ten-year survey \citep[see][]{LSST_numbers_2019}.
    The pixel size is 0.2 arcsec, and the PSF is assumed to be Gaussian with a full width at half maximum size of 0.87 arcsec. 
    
\begin{figure*}[th]
        \centering
        \includegraphics[width=1.0\linewidth]{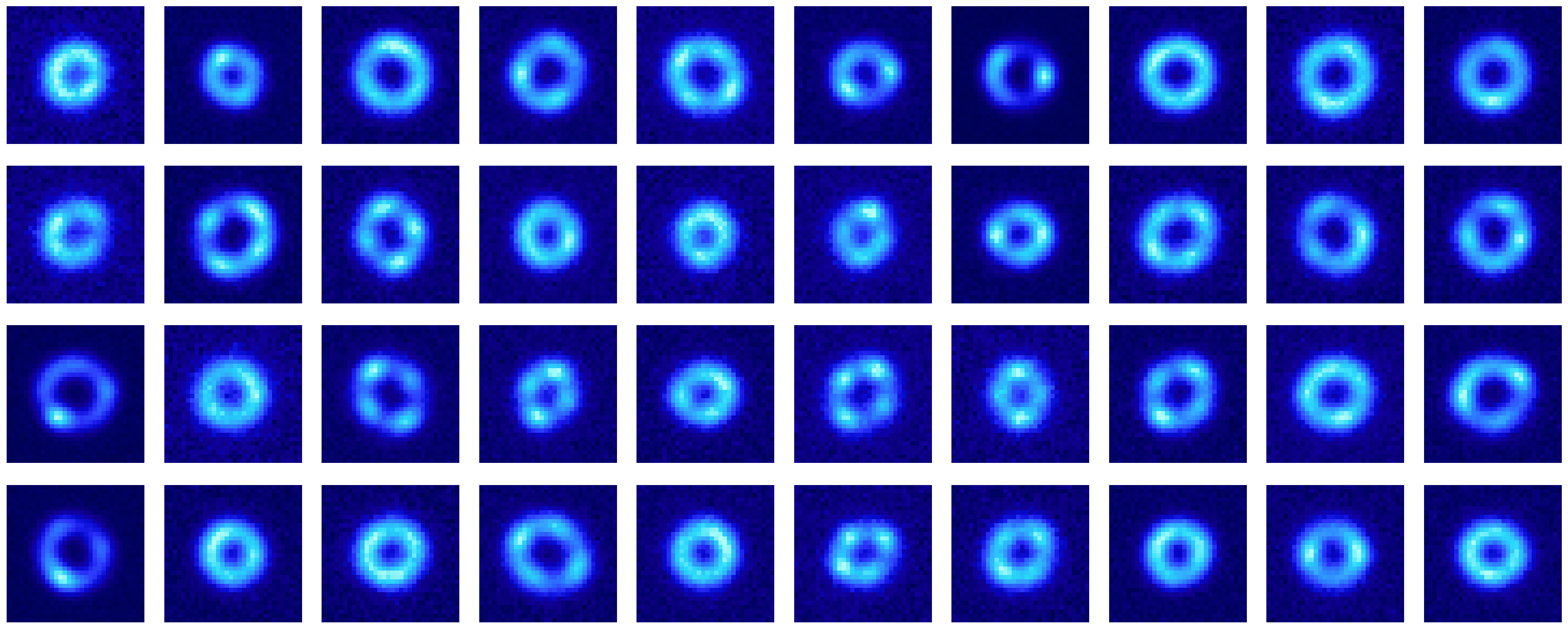}
        \caption{Sample of 40 random strong lenses, generated from the training data distribution for the NRE. The lens systems include a wide variety of closed Einstein rings, doubly imaged sources, and quadruply imaged sources. Such diversity is also seen in real observed lens systems.}
        \label{fig:lenses_caustics}
    \end{figure*}

    Figure~\ref{fig:lenses_caustics} displays 40 randomly sampled strong lens images generated with the training data distribution from Table~\ref{tab:parameter_distribution}. These distributions define the training distribution.
    The figure illustrates the variety of the possible lens configurations that the networks learn to marginalize. The lens systems include Einstein rings, doubles, and quadruples. Such diversity is also observed and expected in real lens systems.
    
    \begin{table}[t]
    \centering
    \caption{Parameter distributions for the simulation of the main deflector and the source light for the training data generation of the NRE. $\mathcal{N}(\mu, \sigma)$ indicates a normal distribution with mean $\mu$ and standard deviation $\sigma$, and $\mathcal{U}[a, b]$ denotes a uniform distribution between $a$ and $b$.}
    \begin{tabular}{l l}
    \hline \hline
         \textbf{Parameter} & \textbf{Distribution} \\ \hline
         \textbf{Lens galaxy} & \\ \hline
         Einstein radius $\theta_{\rm E}$ & $\mathcal{U}[1.0, 1.5]$ \\
         Axis ratio $q_{\rm SIE}$ & $\mathcal{U}[0.5, 0.99]$ \\ 
         Orientation angle $\phi_{\rm SIE}$ & $\mathcal{U}[0.0, \pi]$ \\ 
         Lens center $(\hat{x}_{\rm SIE}, \hat{y}_{\rm SIE})$ & $(0,0)$ \\
         Dark Matter abundance $f_{\rm sub}$ & $\mathcal{U}[0.0, 0.20]$ \\
         DM population mass slope $\beta$ & $\mathcal{U}[-1.5, -0.5]$
         \\ 
        \cr
        \textbf{Source light} & \\ \hline
         Number of sources $N$ & $\mathcal{U}[5-50]$ \\
         Magnitude $mag_{\rm source}$ & $\mathcal{N}(23.5, 0.1)$ \\
         Sérsic index $n_{\rm s\Acute{e}rsic}$ & $\mathcal{N}(2.5, 0.5) \geq 0.8$ \\
         Axis ratio $q_{\rm s\Acute{e}rsic}$ & $\mathcal{U}[0.5, 0.99]$ \\
         Orientation angle $\phi_{\rm s\Acute{e}rsic}$ & $\mathcal{U}[0.0, \pi]$ \\
         Sérsic radius $R_{\rm s\Acute{e}rsic}$ & $\mathcal{N}(0.5, 0.3) \geq 0.05$ \\
         Source center $\hat{x}_{\rm source}, \hat{y}_{\rm source}$ &  $\mathcal{N}(0.0, 0.1)$ \\ \hline \hline
    \end{tabular}
    \label{tab:parameter_distribution}
    \end{table}
    
    \subsection{Training Simulations for the SNPE} \label{sec:train_snpe}
    To simulate the training data of the SNPE, we utilize the built-in strong lensing simulation code in \href{https://github.com/swagnercarena/paltax/}{Platax}\footnote{\url{https://github.com/swagnercarena/paltax/}} by \citet{Wagner_C2023, Wagner_C2024}, which is based on lenstronomy \citep{lenstronomy01, lenstronomy02} and written in JAX. 
    
    The main deflector is modeled with an EPL profile, described in Equation~\ref{eq:EPL_kappa}. 
    The ellipticity of the main deflector is described by two ellipticity parameters $(\gamma_1, \gamma_2)$ given by
    \begin{equation}
    \begin{split}
        \gamma_1 &= \frac{1-q}{1+q} cos(2\phi) \\
        \gamma_2 &= \frac{1-q}{1+q} sin(2\phi)
    \end{split}
    \end{equation}
    with $q$ the axis ratio of the EPL profile and $\phi$ the angle.
    The prior on the ellipticity are in a different space than for the NRE, since the used lensing codes differ in the implementation of the ellipticity for the mass profiles.    

    In the SNPE training distribution we used the in Paltax used tNFW profiles for the individual subhalos, which have, in comparison to the NFW profile, a finite expansion. 
    The subhalo mass function used in the SNPE framework is the same power-law as for the NRE function, but parameterized differently:
    \begin{equation}
        \frac{\text{d}^2 N_{\rm sub}}{\text{d}A \text{d}m_{\rm sub}} = \Sigma_{\rm sub} \frac{m_{\rm sub}^\beta}{m_{\rm pivot, sub}^{\beta +1}}
    \end{equation}
    with $\Sigma_{\rm sub}$ the normalization of the subhalo mass function drawn from a distribution with mean $\Sigma_{\rm sub, pop}$ and standard deviation $\Sigma_{\rm sub, pop, \sigma}$, $\beta$ the mass function slope, and $m_{\rm pivot, sub}$ the pivot mass between the minimal $m_{\rm min, sub}$ and maximal $m_{\rm max, sub}$ rendered subhalo mass. 
    In this case, the slope of the mass function is drawn from a tightly defined uniform distribution $ -2.02 \leq \beta \leq -1.92$, which leaves only the normalization $\Sigma_{\rm sub}$ as a variable parameter to be inferred by the SNPE.

    The background sources are real pixelated galaxy images from a set of 8\,000 elliptical galaxies originating from the DESI survey \citep{Dey_2019_DESI}, selected and compiled in \citet{Missa_2024_Desi_dataset}.
    
\begin{figure*}[ht]
        \centering
        \includegraphics[width=1.0\linewidth]{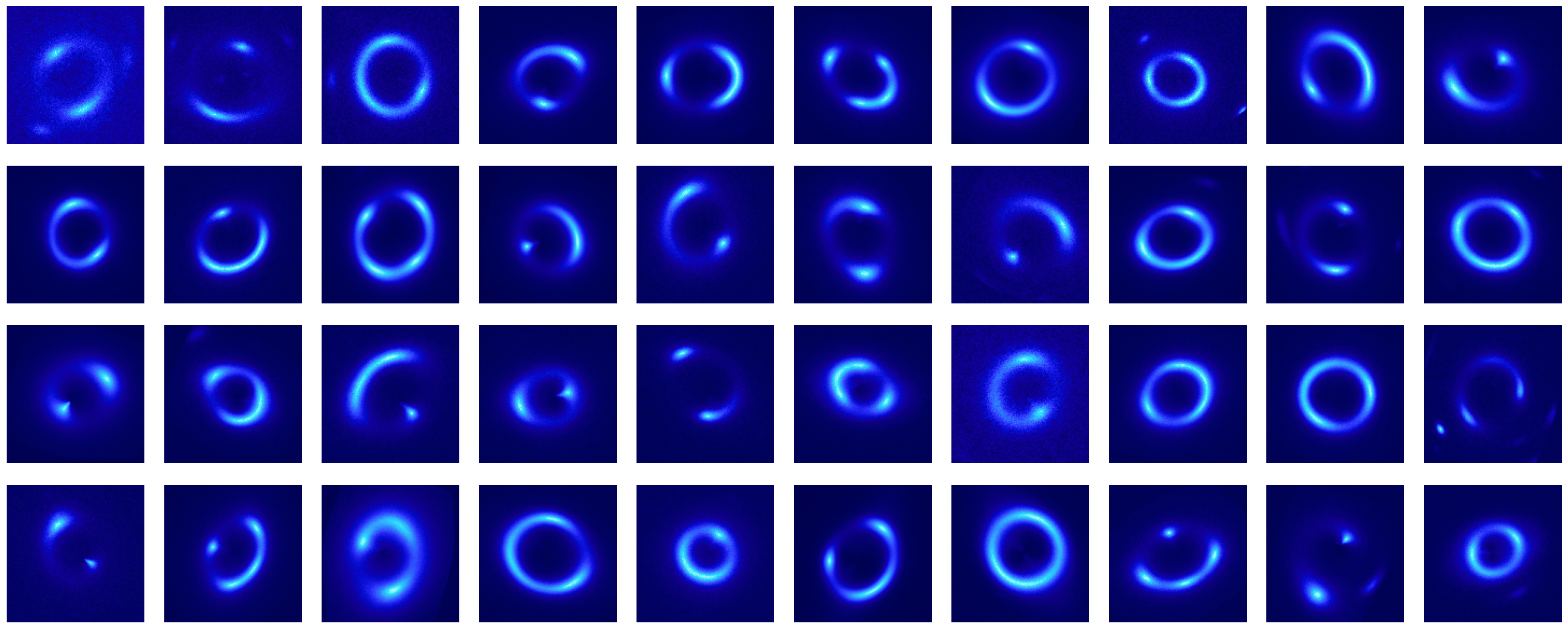}
        \caption{Sample of 40 random strong lenses generated from the training data distribution of the SNPE. The lens systems include a wide variety of closed Einstein rings, doubly imaged sources, and quadruply imaged sources.}
        \label{fig:lenses_NPE}
    \end{figure*}

    Figure~\ref{fig:lenses_NPE} shows a sample of 40 strong lenses from the training data distribution of the SNPE.
    The pixel size is 0.04 arcsec and the PSF is a Gaussian PSF with a full width half maximum size of 0.04 arcsec.
    The observed variety is similar to the images used to train the NRE.
    The pixel size and noise are different from the data used in the NRE, because we use different code bases, and we adapt the NRE image generation to use \texttt{caustics} and data specification to match the expected full ten year LSST survey data quality.
    This can make the sensitivity to systematics more pronounced in the case of the higher resolution data, since sub-pixel changes in the lower resolution of the NRE can be pixel changes in the higher-resolution images of the SNPE.

   The training parameter distribution is displayed in Table~\ref{tab:training_parameter_SNPE}. The table shows the training parameter distributions for the main deflector, for the source images we use -- as mentioned above -- real images.

   \begin{table}[t]
    \centering
    \caption{Parameter distributions for the simulation of the main deflector for the training data generation of the SNPE. $\mathcal{N}(\mu, \sigma)$ indicates a normal distribution with mean $\mu$ and standard deviation $\sigma$.}
    \begin{tabular}{l l}
    \hline \hline
         \textbf{Parameter} & \textbf{Distribution} \\ \hline
         Einstein radius $\theta_{\rm E}$ & $\mathcal{N}[1.1, 0.15]$ \\
         Mass slope $\gamma_{\rm EPL}$ & $\mathcal{N}[2.0, 0.1]$ \\
         Lens center $(\hat{x}_{\rm EPL}, \hat{y}_{\rm EPL})$ & $\mathcal{N}[0.0, 0.16]$ \\
         Ellipticity $(\gamma_1, \gamma_2)$ & $\mathcal{N}[0.0, 0.1]$ \\
         \hline \hline
    \end{tabular}
    \label{tab:training_parameter_SNPE}
    \end{table}

\section{Misspecification Tests}\label{sec:tests}
Real observational data can have distributions that differ from those of training data. Ensuring that the method is insensitive to variations that can occur in real data is crucial for the reliability of the predictions made by machine learning models. \citet{Adversarial_attacks_2013} demonstrated that even minimal, visually imperceptible variations in the input to a network can significantly alter the network's prediction.

We test a number of small variations in the nuisance parameter distributions of the simulations. 
In these tests, we vary one parameter at a time, introducing variations that can be expected in real observations and may be difficult to distinguish from the assumed ground truth distributions of training data.
The modifications are examples of possible, not exact deviations that can occur in real data. We simplify the problem and test only one deviation at once. In reality there can be multiple effects at once that worsen the bias, especially when combining more lenses. We do not include tests for shifts in the parameters of interest, which can further affect the performance of the methods. 

Our OOD and prior misspecification tests vary the following parameters:
\begin{description}
    \item[Source Galaxies] Strong lensing magnifies sources at high redshifts. The statistical distribution of the surface brightness of these high redshift galaxies is not accurately known. The background galaxies in lensing systems are likely to be a biased sample compared to the unlensed galaxy population at the same redshift \citep[e.g.,][]{Yashar_2012}. Moreover, the population-level distribution of the surface brightness of background galaxies is a high-dimensional variable that is difficult to infer from large datasets of lensing systems in a hierarchical framework \citep[although see ][]{Missa_2024_Desi_dataset, Rozet_2024}.  Therefore, it is desirable that minor modifications to this distribution do not bias the inference of dark matter model parameters. We explore this by both changing the distributions of our parametric models (composed of a number of Sérsic components) and also by using pixelated images of galaxies that exhibit different morphologies from those in the training data. 
    \item[Einstein Radius] The existence of a selection bias in the Einstein radii of strong lenses is well understood \citep[e.g.,][]{Mandelbaum_2009, Collett_2016, Sonnenfeld_2023_selectionb}. While we do not change the minimum and maximum Einstein radii in our training data, we modify the distribution of the test data within those bounds. 
    \item[Mass Slope] The mass slope of the main deflector is known to be highly degenerate with the background source morphology \citep[e.g.,][]{Treu_SLACS_2009, Schneider_Sluse_2014}.
    In some experiments, we modify the isothermal density slope by $\sim1\%$. 
    \item[External Shear] The ellipticity of the main deflector is highly degenerate with external shear caused by the environment of the lens \citep[e.g.,][]{Oguri_2005_shear, Koopmans_2006, Sonnenfeld_SL2S_2013, Shu_2016, Shu_2017, Talbot_2021_eBOSS}. We introduce external shear of $\sim 0.1$ to test data in experiments with no shear in training data.
    \item[Subhalo Profiles] The true profiles of dark matter subhalos are unknown \citep[e.g.,][]{Heinze_2024_NFW_dont_fit_tng50}. We explore the effect of changing NFW profiles to truncated NFW profiles and vice versa to quantify the bias introduced in the inferred mass function.    
    \item[Observational Noise] The distribution of observational noise depends on the instrument, the time and duration of observations, and many other variables. Without introducing non-Gaussianity to the noise, we study the effect of increasing the variance of the noise in the test data.
\end{description}

Although for real data all the listed variations can occur simultaneously, in this case study, we apply one effect at a time to isolate the effects of individual variations. 
The specific modified parameters to test the NRE and the SNPE frameworks can be found in Table~\ref{tab:mods} and~\ref{tab:mods_SNPE_tests} respectively.

\section{Results and Discussion}
\label{sec:results}
This paper aims to show that even minimal variations in the underlying data generation distribution affects the neural network predictions and can bias these in unpredictable and therefore uncorrectable ways. 

Under real observing conditions any modification compared to the training data generation can occur at any strength. Therefore, any biased caused by any modification shows the fragility of the neural network predictions.
The shown modifications are examples of possible, not exact deviations that can occur in real data. We simplify the problem and also test each single deviation at once. In reality there can be multiple effects at once that worsen the bias, especially when combining more lenses.

We aim to show these biases quantitatively and only determine in which $\sigma$ interval the ground truth lies under different modifications in the data generating process. 
We show that the biases are pronounced when combining a large number of lenses and dominate the inference. Additionally, we provide a per lens quantitative assessment using TARP \citep{Lemos_2023_TARP} in appendix~\ref{app:TARP} to show that not only the population-level inference, but also the per-lens inference is biased for an OOD modification.

    \subsection{Neural Ratio Estimator}
\begin{figure*}[htbp]
        \centering
        \begin{minipage}[t]{0.45\textwidth}
            \centering
            \includegraphics[width=\textwidth]{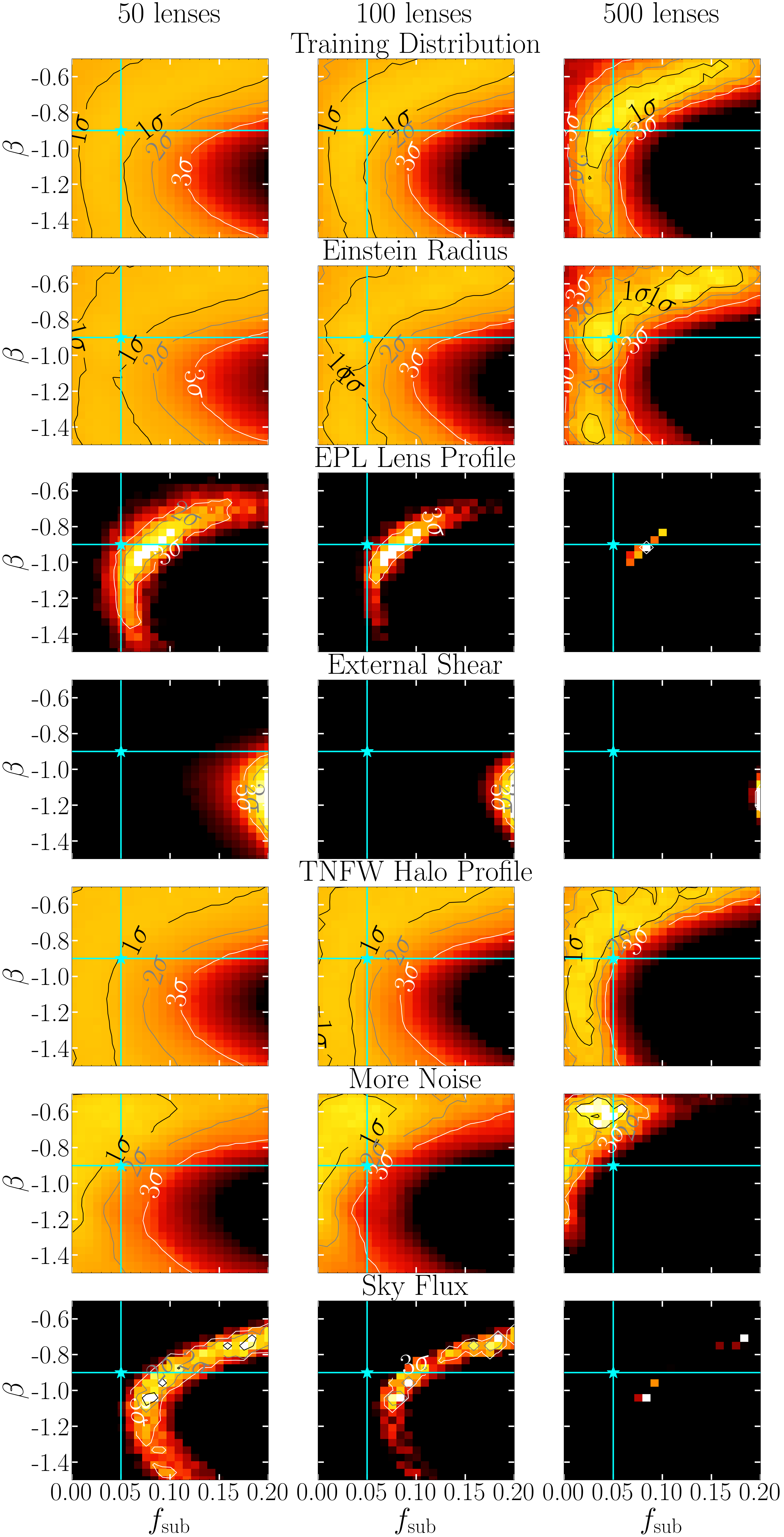}
        \end{minipage}%
        \begin{minipage}[t]{0.45\textwidth}
            \centering
            \includegraphics[width=\textwidth]{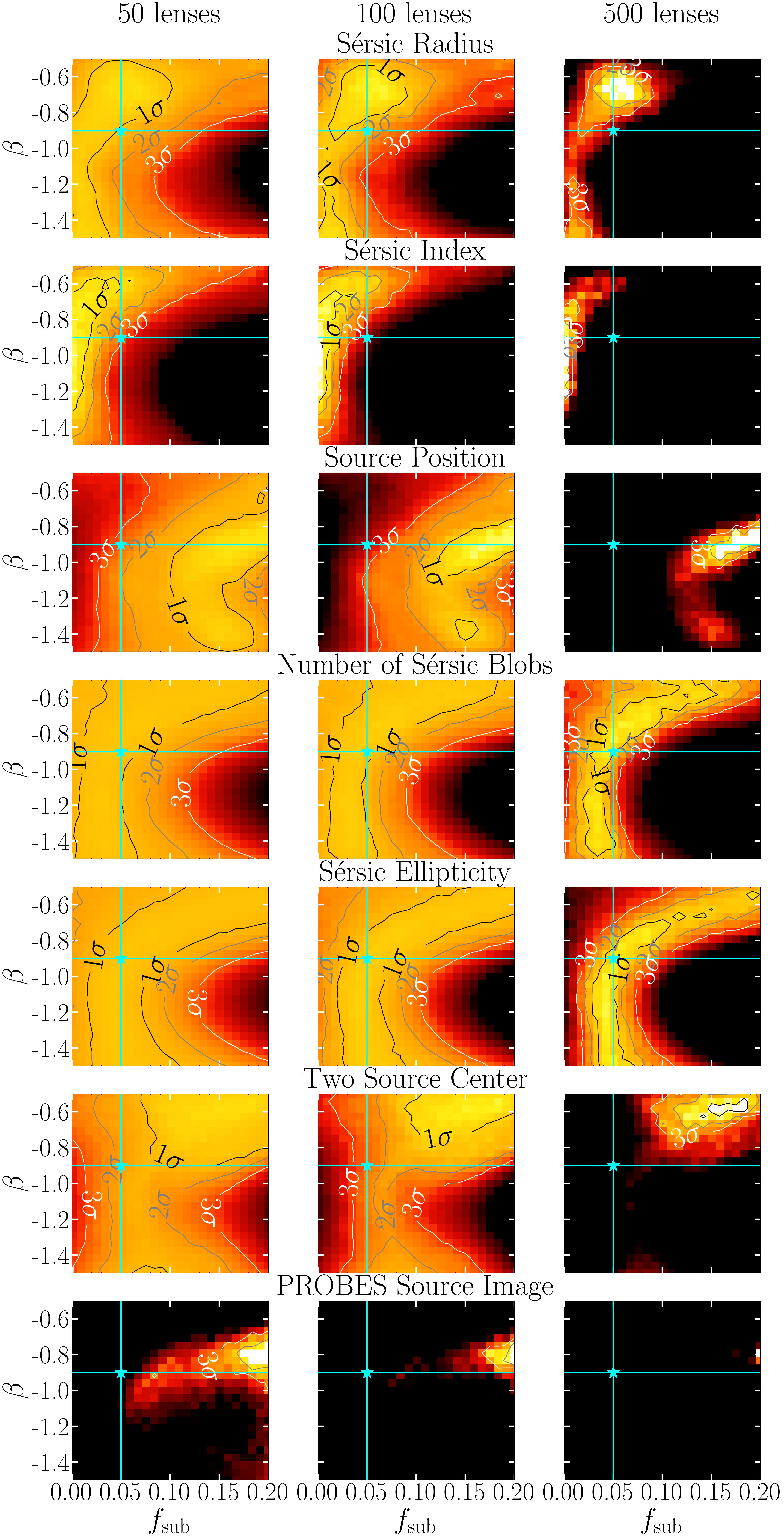}
        \end{minipage}%
        \begin{minipage}[t]{0.08\textwidth}
            \centering
            \raisebox{11mm}{
            \includegraphics[width=\textwidth]{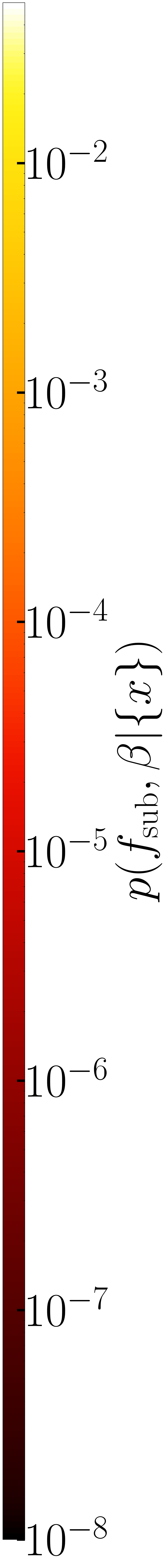}
            }
        \end{minipage}%
        
        \caption{The evaluation of the NRE is conducted on lenses drawn from the training distribution and from distributions with minor variations in the parameters, following Section~\ref{sec:tests}. 
        All evaluation datasets were generated with the ground truth parameters set at $f_{\rm sub} = 0.05$ and $\beta = -0.9$, which are in accordance with $\Lambda$CDM predictions and are marked by a cyan star. The modifications to the parameter distributions for the individual plots are detailed in Table~\ref{tab:mods}.}
        \label{fig:variations}
    \end{figure*}

    We evaluate the calibrated NRE on lenses that are drawn from the training distribution as well as from distributions with minor variations as defined in Table~\ref{tab:mods}, following the tests in Section~\ref{sec:tests}.
    The inferred posteriors from the original and the varied distributions are shown in Figure~\ref{fig:variations}. 
    All images used for the posterior inference were generated with the ground truth parameters set at $f_{\rm sub} = 0.05$ and $\beta = -0.9$, which are consistent with $\Lambda$CDM predictions \citep[e.g.,][]{Diemand_2007_ViaLactea, Kuhlen_2007_ViaLactea, Springel_2008, Hiroshima_2018}. The ground truth is denoted by the cyan star in the posterior predictions. In all experiments, we only modify the distribution of one parameter at a time, with the remaining parameters sampled from the same distribution as the training set (Table~\ref{tab:parameter_distribution}).
    The prior on the subhalo abundance and the mass slope are, as defined in Table~\ref{tab:parameter_distribution}, uniform distributions with $f_{\rm sub} \in \mathcal{U}[0.0, 0.20]$ and $\beta \in \mathcal{U}[-1.5, -0.5]$.

    \begin{table*}[t]
        \centering
        \caption{Parameter distribution variations used to create Figure~\ref{fig:variations}. The modifications are described in more detail in Section~\ref{sec:tests}. Here, $\mathcal{N}(\mu, \sigma)$ represents a normal distribution with mean $\mu$ and standard deviation $\sigma$, and $\mathcal{U}[a, b]$ denotes a uniform distribution between $a$ and $b$.}
        \begin{tabular}{l  l  l }
        \hline \hline
             \textbf{Plot Title} & \textbf{Modified Parameter} & \textbf{New Distribution} \\ \hline 
             Training Distribution & None & \\ \hline
             Einstein Radius & Einstein radius $\theta_E$ & $\mathcal{N}(1.4, 0.2) \geq1.0, \leq1.5$  \\ \hline
             EPL Lens Profile & Power-law slope of lens $\gamma_{\rm EPL}$ & $\mathcal{N}(2.0, 0.02)$ \\ \hline
             External Shear & Added external shear components $\gamma_{1/2}$ & $\mathcal{N}(0.0, 0.1)$ \\ \hline 
             TNFW Halo Profile & Truncation of NFW, truncation scale $\tau$ & $5$ \\ \hline 
             More Noise & Noise in the image at 1\% of the sky flux & \\ \hline
             Sky Flux & Different magnitude of sky background $mag_{\rm sky}$ & $21.4$ \\ \hline 
             Sérsic Radius & Sérsic radius $R_{\rm s\Acute{e}rsic}$ & $\mathcal{N}(0.2,0.2) \geq 0.05$ \\ \hline
             Sérsic Index & Sérsic index $n_{\rm s\Acute{e}rsic}$ & $\mathcal{N}(5.5, 1.5)\geq0.8$ \\ \hline
             Source Position & Source centers $(\hat{x}_{\rm source}, \hat{y}_{\rm source})$ & $\mathcal{N}(0.05, 0.15)$ \\ \hline 
             Number of Sérsic Blobs & Number of Sérsic blobs in the source $N$ & $\mathcal{U}[40-45]$ \\ \hline
             Sérsic Ellipticity & Source axis ratio $q_{\rm s\Acute{e}rsic}$ & $\mathcal{N}(0.75, 0.2) \geq 0.5, \leq0.99$ \\ \hline
             Two Source Center & Source position $(\hat{x}_{\rm source 1}, \hat{y}_{\rm source 1}), (\hat{x}_{\rm source 2}, \hat{y}_{\rm source 2})$ & $\mathcal{N}_1(0.0, 0.1), \mathcal{N}_2(-0.2, 0.1)$ \\ \hline
             Probes Dataset as Sources & Analytic Sérsic source profiles to Probes dataset & \\ \hline \hline
        \end{tabular}
        \label{tab:mods}
    \end{table*}

    \begin{table*}[t!]
        \centering
        \caption{Parameter distributions variations used to create Figure~\ref{fig:NPE_results}. The modifications are described in more detail in Section~\ref{sec:tests}. $\mathcal{N}(\mu, \sigma)$ represents a normal distribution with mean $\mu$ and standard deviation $\sigma$.}
        \begin{tabular}{l  l  l }
        \hline \hline
              & \textbf{Modified Parameter} & \textbf{New Distribution} \\ \hline 
             Training Distribution & None & \\ \hline
             Bigger Einstein Radius & Einstein radius $\theta_E$ & $\mathcal{N}(1.5, 0.1)$  \\ \hline
             Smaller Einstein Radius & Einstein radius $\theta_E$ & $\mathcal{N}(0.7, 0.1)$ \\ \hline 
             NFW Halo Profiles & Usage of NFW mass profile for subhalos &  \\ \hline 
             Spiral Source Galaxies & SKIRT TNG dataset as source images & \\ \hline
             Added Blobs & Addition of Sérsic light blobs to sources & 30-50 Sérsic profiles \\ \hline             
             AstroCLIP split & Elliptical galaxies are split with AstroCLIP & \\ \hline \hline
        \end{tabular}
        \label{tab:mods_SNPE_tests}
    \end{table*}

    Figure~\ref{fig:variations} is organized as follows: 
    The left-hand side displays variations in the lens plane and noise levels, while the right-hand side shows variations in the source. The columns correspond to the posteriors inferred by combining 50, 100, and 500 lens systems, each row illustrating the effect of a different test.
    
    The posterior inferred from lenses generated within the training distribution is shown in the top row of the left-hand side. The inferred posterior is unbiased and recovers the ground truth well, indicating that even inference with a large number of lensing systems remains unbiased.
    
    In the second row on the left, we vary the Einstein radius distribution from uniform to a normal distribution with mean $\mu = 1.4$ and standard deviation $\sigma = 0.2$, while maintaining the same boundaries as in the training distribution. 
    Although the posteriors for 50 and 100 lenses still recover the ground truth, the posterior for 500 systems develops multiple modes. This reflects the mismatch between the training and testing distributions, an effect that becomes more pronounced with a larger sample and can be detected -- if the ground truth is known -- using coverage metrics such as TARP \citep{Lemos_2023_TARP}.
    
    The third row shows the posterior when changing the main deflector model from a SIE to an EPL profile, with the mass slope drawn from a normal distribution with mean $\mu = 2.0$ and standard deviation $\sigma = 0.02$. For a slope of $\gamma_{EPL} = 2.0$, the EPL profile corresponds to a SIE profile. Despite this minimal variation ($\sim1\%$), even for 50 lenses the posterior is highly  biased, and the ground truth is outside the $3\sigma$ region. When combining 500 lenses, the probability density becomes highly concentrated at a very narrow value, which can be due to network failure under this specific modification.
    
    In the fourth row, we introduce a small external shear defined by two shear components $\gamma_{1,2}$ each with a mean of $\mu = 0.0$ and standard deviation $\sigma = 0.1$. 
    The shear is highly degenerate with the ellipticity of the lens profile and is typically used to model the local lens environment \citep[e.g.,][]{Oguri_2005_shear, Koopmans_2006, Sonnenfeld_SL2S_2013, Shu_2016, Shu_2017, Talbot_2021_eBOSS}. The additional external shear significantly biases the network output, pushing the obtained posterior towards and out of the priors on the parameters of interest which are the same as the priors used for training data generation, defined in Table~\ref{tab:parameter_distribution}. The ground truth is already outside the $3\sigma$ region for only 50 combined lenses.
   
    The fifth row introduces a modification to the dark matter halo mass profiles, changing from NFW to truncated NFW profiles with a truncation radius five times the scale radius, converting infinitely extended NFW profiles to finite mass profiles. The resulting posterior exhibits a clear bias. When combining 50 or 100 lenses, the ground truth is between $1\sigma$ and $2\sigma$, and at 500 lenses already between $2\sigma$ and $3\sigma$.
    
    The sixth row examines the effects of increasing the variance of the observational Gaussian noise by $\sim1\%$ of the sky flux (the zero point of the image), which significantly biases the networks output. The ground truth is recovered within the $2\sigma$ interval for the combination of 50 lenses, but clearly outside the $3\sigma$ interval for 500 lenses.
    The last row on the left side shows the impact of reducing the image zero point. The posterior is strongly biased, highlighting the sensitivity to variations in sky flux and noise distributions.
    For the modification of the sky flux the probability density becomes -- similar to the profile change to an EPL -- highly concentrated at a very narrow value, which can be due to network failure under this specific modification. The ground truth is not even within $3\sigma$ for as few as 50 lenses.

    The right panels of Figure~\ref{fig:variations} show the posteriors obtained when modifying the distribution of the background source parameters. Rows one to six correspond to variations in size, slope, position, complexity (number of Sérsic components), source ellipticity, and number of primary sources behind a deflector. Almost every modification results in a strong bias when combining multiple lenses. The only modifications that include the ground truth within the $3\sigma$ interval when combining 500 lenses are the modifications of the number of Sérsic background sources and the modifications of the background source ellipticity.
    However, the strongest bias is observed in the last row, exploring the effect of using real images of galaxies from the PROBES dataset \citet{Connor_2019_Probes} -- a catalog of over 2500 spiral galaxies from six deep imaging and spectroscopic surveys -- for testing the model.

    These tests reveal that the bias introduced to the posteriors is highly unpredictable and therefore uncorrectable. In some cases, shifts in parameters that significantly impact the images, such as the size of the background sources, result in relatively minor biases. Conversely, changes in parameters that are highly degenerate with others, like external shear and ellipticity, can induce substantial biases in the posteriors, even when they cause nearly imperceptible changes in the data.
 
    The initial parameterization used to train the NRE is within the range of usual state-of-the-art parameterization for lens modeling, and the deviations tested here are mostly physically very small and are very likely to occur in a realistic data analysis setting where the ground truth is not available. 
    The effects of the changes in the underlying parameter distributions on the NRE predictions are highly nonlinear and unpredictable, which makes it difficult to trust inferred posteriors with this model. This makes NREs applicable to cases where both, the underlying distribution of parameters and the physical model, are perfectly known (e.g., for sensitivity predictions). For strong lensing specifically and astrophysical data analysis in general, this is very rarely the case.

    \subsection{Neural Posterior Estimator}    
    By incorporating a hierarchical inference step, which effectively models the prior distribution, the SNPE framework can correct for distributional shifts on parameters included in the population-level model. In our case these are the dark matter population parameters.
    This allows for accurate inference of the parameters of interest even with misspecified priors, provided these priors remain within the test distribution and are low dimensional. 
    
    We first test the SNPE for the scenario of a parameter that is not included in the population-level model -- the Einstein radii -- is sampled from a different distribution, but still within the bounds of the training data.
   
    We then empirically test the robustness of SNPE to data generated outside the bounds of the training data. We do this by considering modifications to the dark matter subhalo profiles (replacing tNFW profiles with NFW profiles) and also changing the distribution of the background source morphologies as detailed in Table~\ref{tab:mods_SNPE_tests}.
    For the population-level inference of the dark matter model we use ten mock targets for each test.
    This is significantly less than for the NRE tests. Each single lens systems requires a significant amount of compute, since each time a new set of sequential neural networks needs to be trained.

    To account for this, the target images for the SNPE are generated using the same random seed, ensuring that, between the evaluations on the training distribution and the modifications, only the varied parameter changes. 
    We do this, to be able to be able to qualitatively see the effects of the conducted tests without the need for a big sample size because of computational limitations. Since we did not experience such high computational costs for tests on the NRE, we did not fix the random seed in the NRE specific tests.
    This implies that the specific realization of positions and masses of the subhalos, along with all other unmodified parameters (including the specific realization of observational noise), remain consistent across the ten different evaluated images. 
    The only exception to this is the variation of the Einstein radius, since our parameterization of the abundance  of subhalos relies on the lens mass, which is directly correlated to the Einstein radius.
    We expect that for the inference to be robust against these modifications, the posterior of the dark matter normalization, $\Sigma_{\rm sub}$, should remain consistent across the various tests.

\begin{figure*}[htbp]
        \centering
        \begin{minipage}[t]{0.49\textwidth}
            \centering
            \includegraphics[width=\textwidth]{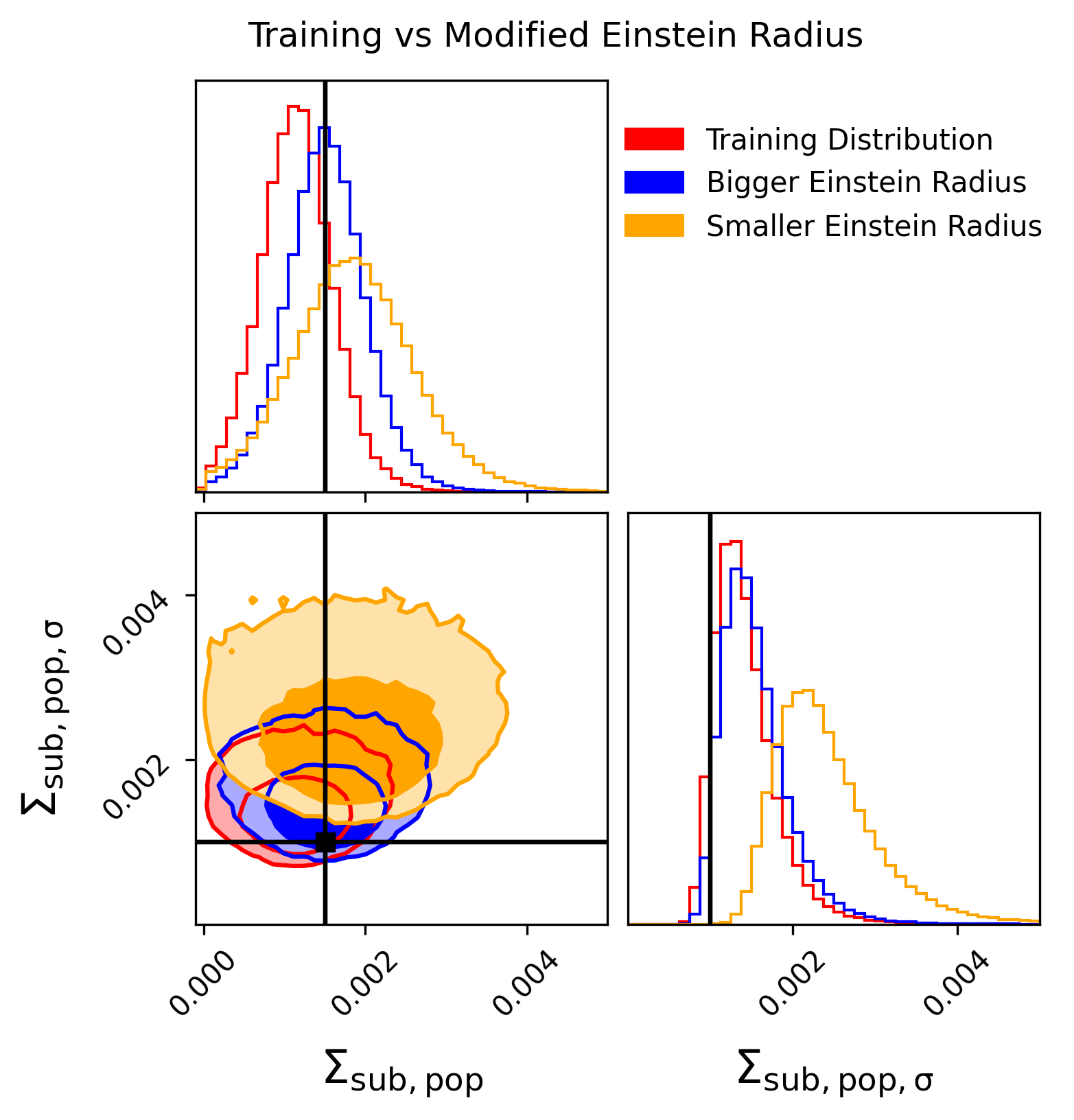}
        \end{minipage}%
        \begin{minipage}[t]{0.49\textwidth}
            \centering
            \includegraphics[width=\textwidth]{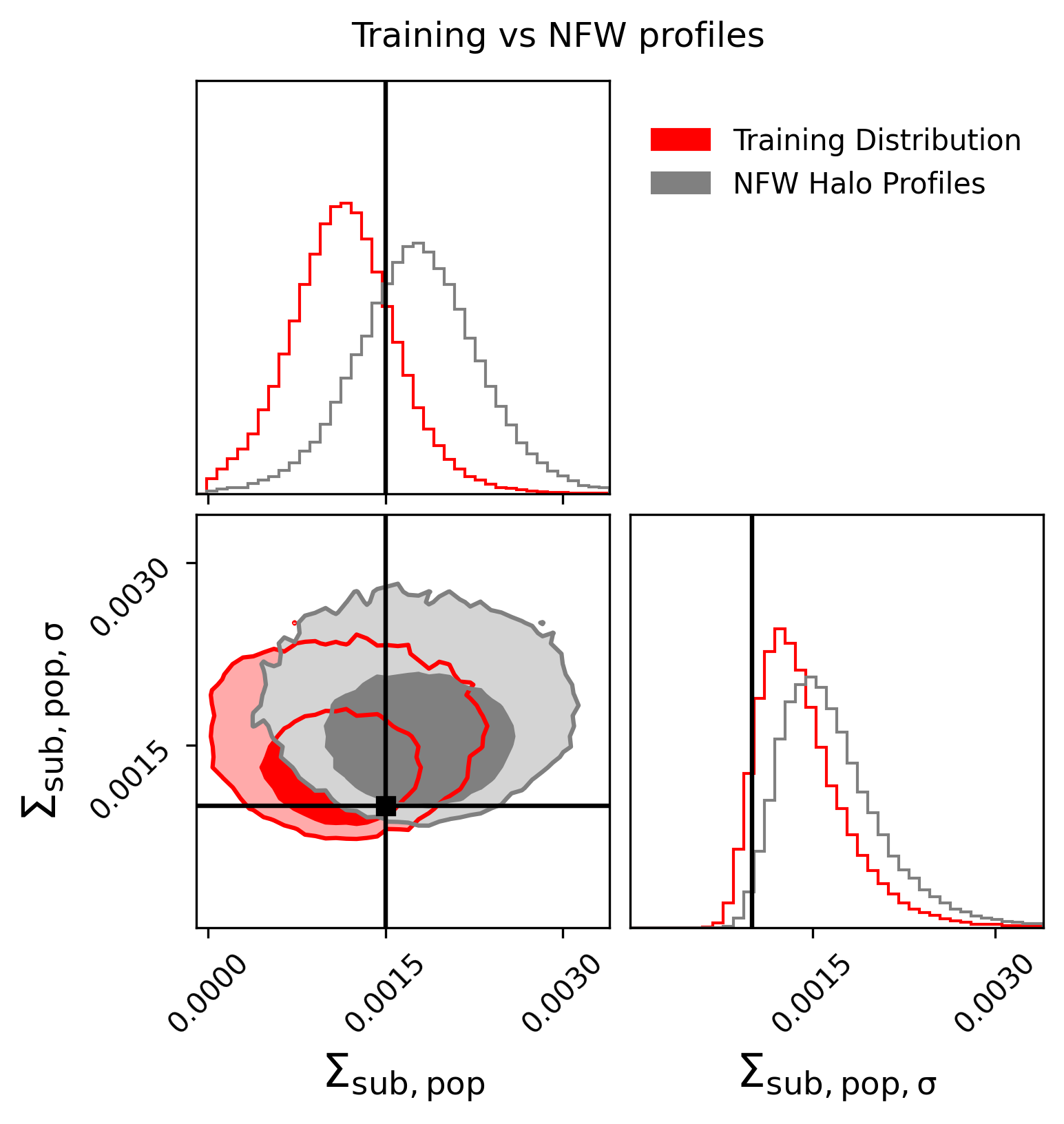}
        \end{minipage}
        
        \begin{minipage}[t]{0.49\textwidth}
            \centering
            \includegraphics[width=\textwidth]{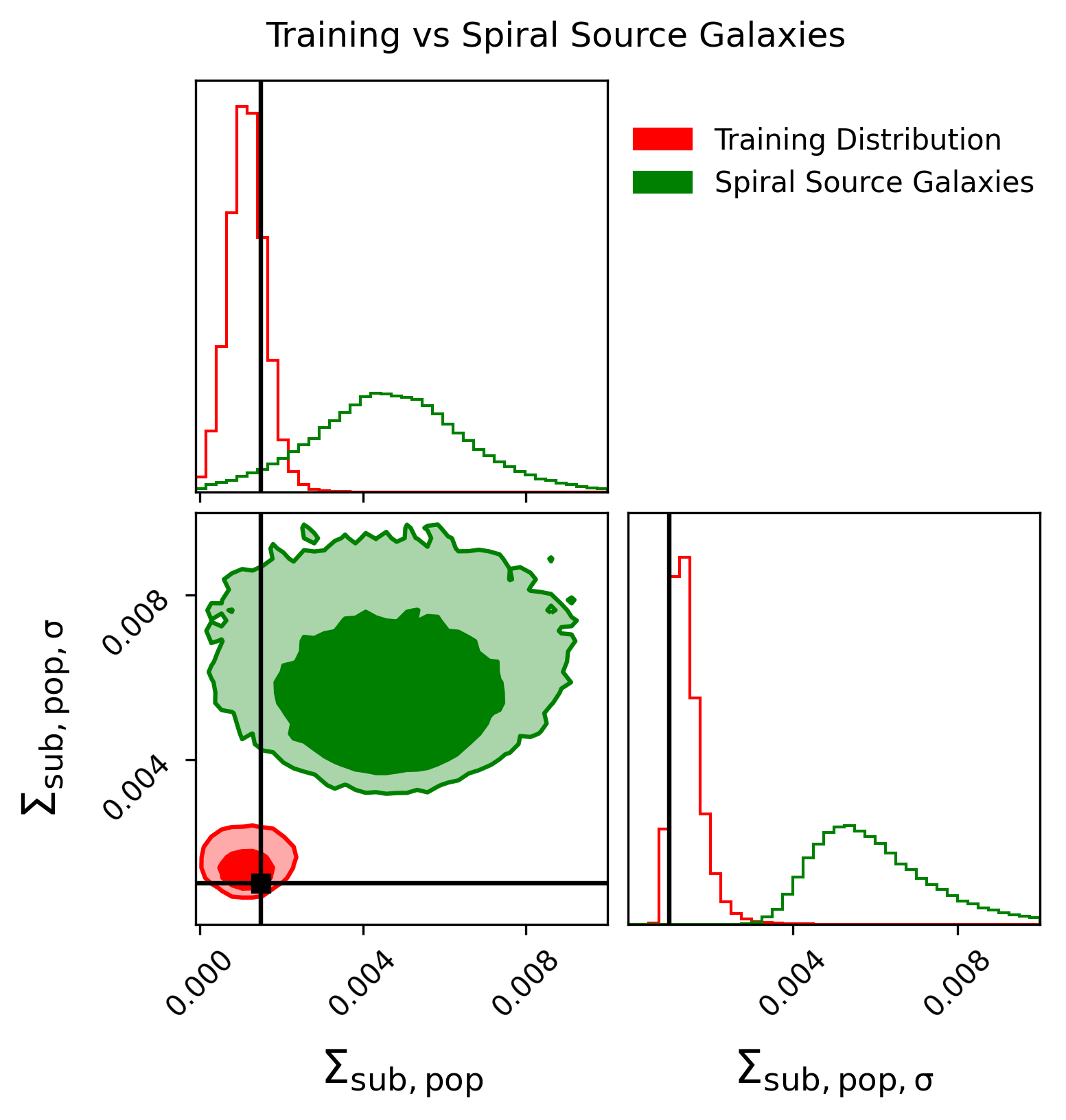}
        \end{minipage}%
        \begin{minipage}[t]{0.49\textwidth}
            \centering
            \includegraphics[width=\textwidth]{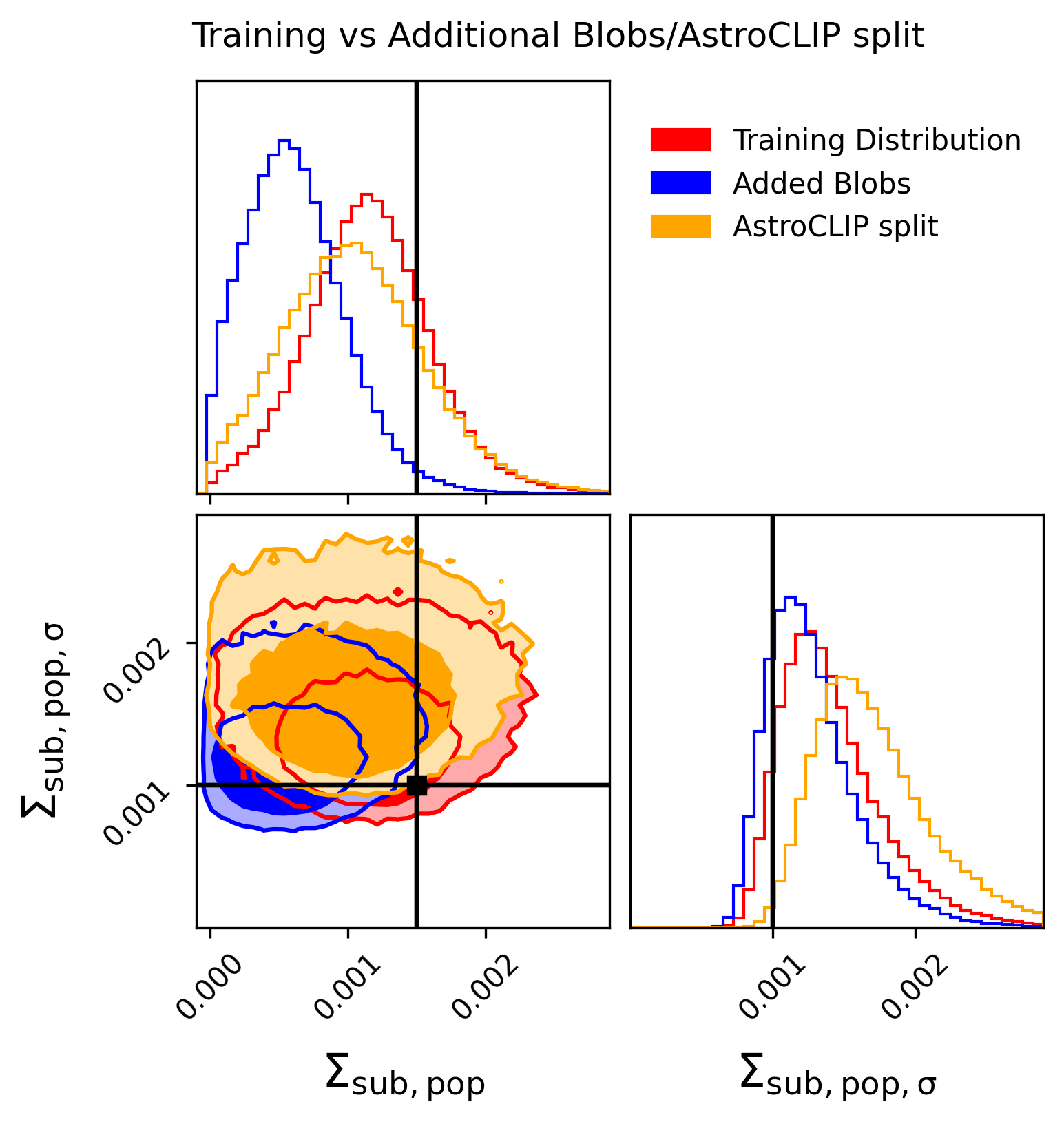}
        \end{minipage}
    
        \caption{Evaluation of the SNPEs on lenses drawn from the training distribution and on variations in the underlying parameter distributions, following the tests described in Section~\ref{sec:tests}. We modify the Einstein radius distribution, the subhalo mass profile, and the source profiles. Details can be found in Table~\ref{tab:mods_SNPE_tests}. The red contours always show the SNPE results for inference on the training data distribution.}
        \label{fig:NPE_results}
    \end{figure*}

    Figure~\ref{fig:NPE_results} illustrates the effect of different tests. 
    The red contours show the SNPE results for inference on the training data distribution, while the other colors show the effects of different modifications (Table~\ref{tab:mods_SNPE_tests}), based on the tests from Section~\ref{sec:tests}. The conducted tests for the SNPE are fewer than for the NRE, since each single lens systems for each individual test requires a significant amount of compute, as each time a new set of sequential neural networks needs to be trained.

    It is important to note that the prior on the subhalo abundance for the training distribution, used to generate the training data, is a narrow normal distribution with mean $\Sigma_{\rm sub, pop} = 0.002$ and standard deviation $\Sigma_{\rm sub, pop, \sigma} = 0.001$. 
    Our target distribution exhibits a shift to mean $\Sigma_{\rm sub, pop} = 0.0015$ and standard deviation $\Sigma_{\rm sub, pop, \sigma} = 0.001$, which can and is corrected for by the hierarchical inference step.    
    
    As seen in Figure~\ref{fig:NPE_results}, this narrow prior limits the range of biases. If the prior is broadened to either a uniform distribution or a normal distribution with a broader standard deviation, the observed biases are expected to increase.

    The specific modifications used in each test from Figure~\ref{fig:NPE_results} are detailed in Table~\ref{tab:mods_SNPE_tests}.
    The top left panel in Figure~\ref{fig:NPE_results} shows the effect of modifying the Einstein radius distribution. The training distribution draws the Einstein radius from a normal distribution with mean $\mu = 1.1$\,arcsec and standard deviation $\sigma=0.15$. The yellow contours show the effect of drawing the Einstein radius from a normal distribution with $(\mu_{\rm small} = 0.7, \sigma_{\rm small}=0.1)$, and the blue contours correspond to drawing the Einstein radius from a normal distribution with $(\mu_{\rm big} = 1.5, \sigma_{\rm big}=0.1)$. 
    This modification demonstrates that lens systems with larger Einstein radii are more constraining on the dark matter abundance than systems with smaller radii. 
    The results suggest that a modification of the Einstein radius distribution does not bias the inferred posteriors within this framework, but in general, systems with larger Einstein radii are more constraining than systems with smaller Einstein radii. 
    Whilst the mean of the subhalo population parameter $\Sigma_{\rm sub, pop}$ is recovered for all cases within $1\sigma$, the predicted standard deviation $\Sigma_{\rm sub, pop, \sigma}$ for the case of generally smaller Einstein radii is outside the $3\sigma$ area. 

    The second variation, shown in the top right panel of Figure~\ref{fig:NPE_results}, changes the subhalo profiles from truncated NFW profiles to NFW mass profiles. The gray contours show the effect on the inferred dark matter normalization constant. The bias in the posterior estimation resulting from this minor change suggests that the specific density profile used in training simulations can negatively impact the accuracy of the inference on the mass function parameters. Note that even within CDM cosmologies, there is theoretical uncertainty in the expected profiles of low-mass subhalos \citep[e.g.,][]{Heinze_2024_NFW_dont_fit_tng50}.
    The ground truth lies between the $1\sigma$ and $2\sigma$ contours when the tNFW halo profile is replaced by an NFW profile. Since the random seed is the same between the training distribution and the test set, there should be no difference between the predictions of the network. Any small deviation can grow drastically and unpredictably when combining a larger number of lenses than ten for inference. 

    The next variations, in the bottom left panel, test the effect of background sources that are outside of the training distribution. For the training data, we used elliptical source images from the DESI survey \citep{Dey_2019_DESI}, selected by \citet{Missa_2024_Desi_dataset}. The green contours originate from target images generated using the SKIRT TNG dataset \citep{Bottrell_2024_SKIRT}, simulated for the Hyper Suprime-Cam Subaru Strategic Program \citep{HSC_whitepaper}. Clearly the ground truth is not within the $2\sigma$ contours. This dataset contains only spiral galaxies. The strong bias observed highlights that precise knowledge of the source profile is required for unbiased dark matter inference.

    Since the difference between spiral and elliptical galaxies is significant, so we also explored the performance of the SNPE network under more subtle source variations, shown on the bottom right panel of Figure~\ref{fig:NPE_results}.

\begin{figure}[t]
        \centering
        \includegraphics[width=0.49\textwidth]{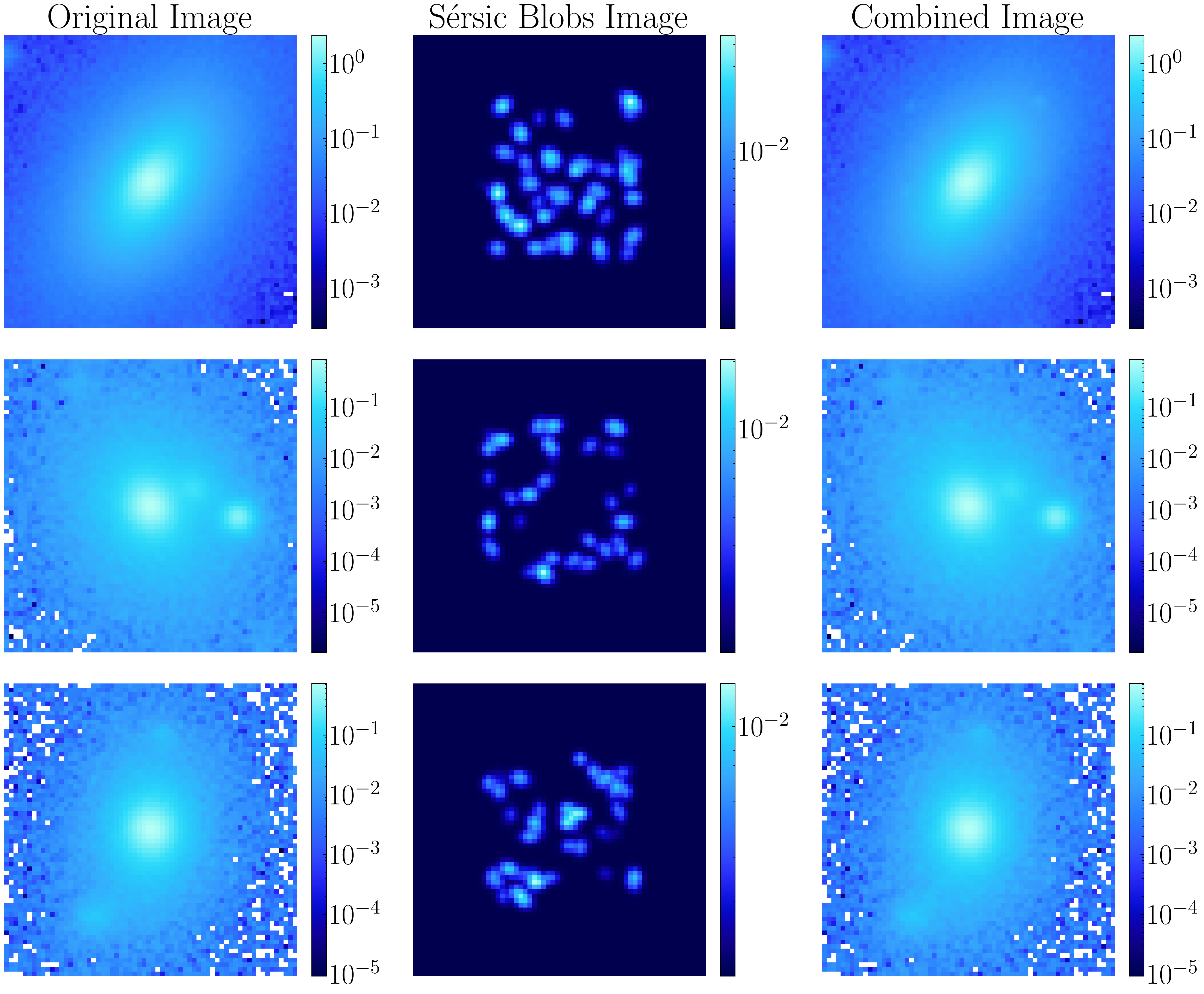}

        \caption{Examples of Sérsic blobs added to the training source image to create OOD data. The differences are barely visible to the eye.}
        \label{fig:Blobs}
    \end{figure}

    First, we took the original elliptical source galaxy images and added between 30 and 50 small Sérsic light profiles, with a brightness of at least an order of magnitude less than the main background source galaxy. The bottom right panel of Figure~\ref{fig:Blobs} shows examples of these variations. By eye, these changes are barely visible. 
    The motivation behind this modification is that, since the true distribution of lensed high-redshift galaxies is unknown, they could exhibit more or less structure than samples of unlensed galaxies at comparable redshifts.
    The effect of these modifications on the SNPE prediction is shown in the blue contours, indicating that the true distribution is now outside the $2\sigma$ contours of the posterior on the variational parameters, which will get even more pronounced when doing inference on more strong lenses.
    
\begin{figure*}[htbp]
        \centering
        \begin{minipage}{0.05\textwidth}
            \rotatebox{90}{
            \begin{tabular}{c} 
                AstroCLIP \\  
                Group 1
            \end{tabular}
        }
        \end{minipage}%
        \begin{minipage}{0.94\textwidth}
            \includegraphics[width=\linewidth]{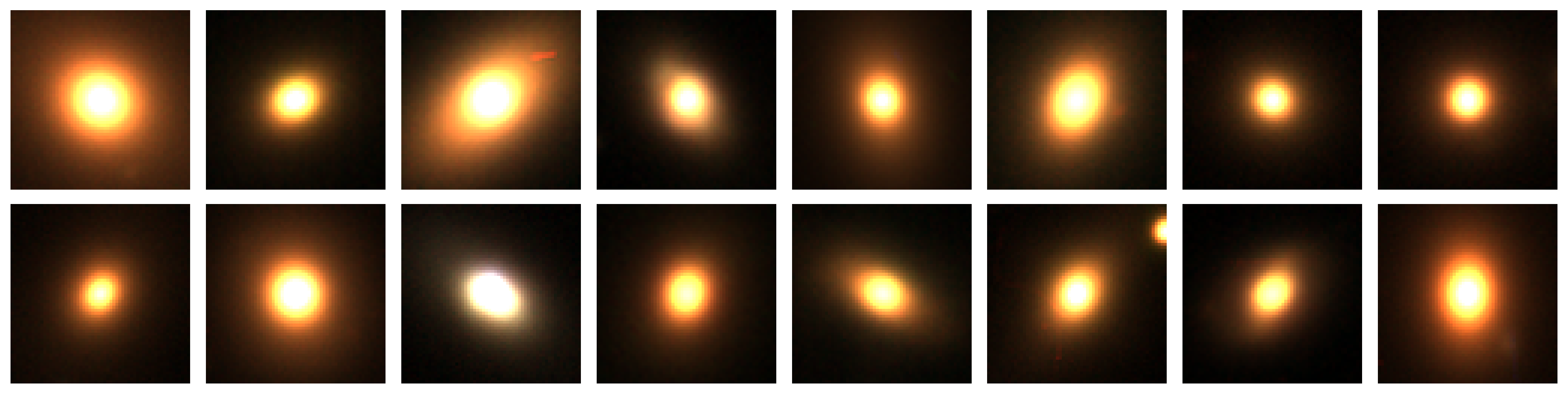}
        \end{minipage}%
        \vspace{10pt} 
        \begin{minipage}{0.05\textwidth}
            \rotatebox{90}{
            \begin{tabular}{c}  
                AstroCLIP \\ 
                Group 2
            \end{tabular}
        }
        \end{minipage}%
        \begin{minipage}{0.94\textwidth}
            \includegraphics[width=\linewidth]{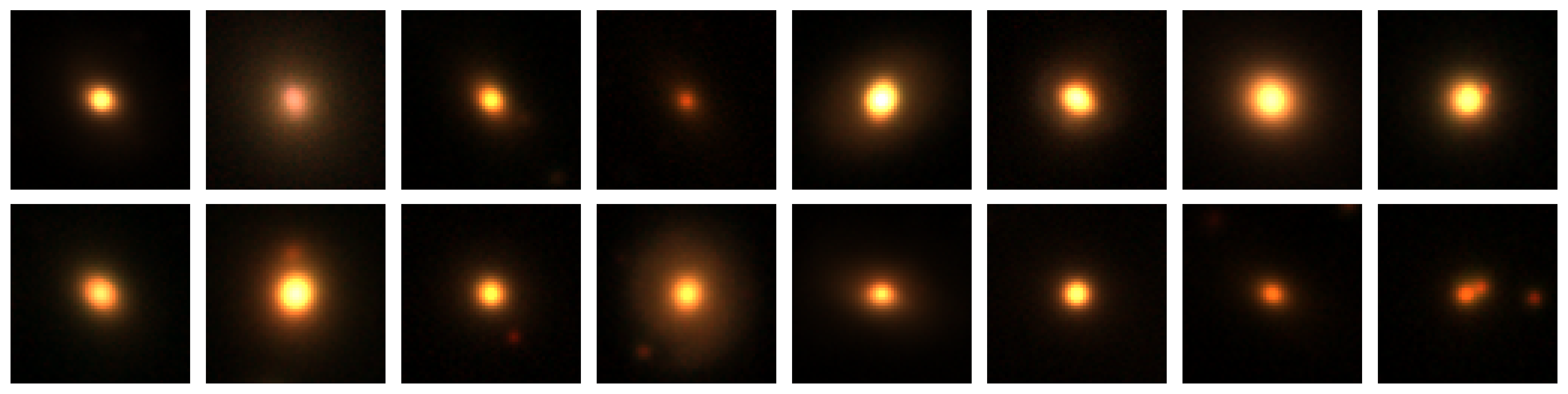}
        \end{minipage}%
        \caption{Examples from the two groups of elliptical DESI images grouped using AstroCLIP \citep{AstroCLIP_2024}. The first group is used to generate the target images, and the second group is used for training data generation.}
        \label{fig:AstroCLIP}
    \end{figure*}

    As an additional test, we split the elliptical galaxies into two datasets using AstroCLIP \citep{AstroCLIP_2024}, a foundation model trained on real galaxy images, to embed the dataset into a latent space. We then split the galaxies into two groups using the K-means clustering algorithm implemented in scikit-learn\footnote{\url{https://scikit-learn.org/stable/modules/clustering.html\#k-means}}. Examples from the two groups are shown in Figure~\ref{fig:AstroCLIP}. 
    The first group is used to generate the target image, and the second group is used for training data generation. Although we observe a shift in the posterior, the bias is less significant compared to the added blob test. We speculate that this is due to the fact that this specific clustering of the data is primarily separating the background galaxies based on their size (Figure~\ref{fig:AstroCLIP}) while the two classes still have similar small scale surface brightness fluctuations, which is known to be a key factor for detecting subhalos \citep[e.g.,][]{Yashar_2013}. 
    The ground truth lies on the $2\sigma$ contour, indicating a small, but non-negligible shift to the training distribution.
    
    As in the case of the NRE, these tests show that the SNPE framework is prone to specific biases when testing on data with a different distribution compared to the training data, specifically in the high dimensional space of the background sources, which are known to have strong degeneracies with the subhalo parameters \citep[e.g.,][]{Yashar_2013, Vegetti_2014, Vegetti_2023}.
    These test do not aim to qualify any shifts in terms of severity of effects, but rather show that any minor modification that can be well expected in real observed data can lead to unpredictable biases that are not easy to correct for.

\section{Conclusion}
\label{sec:Conclusion}
This paper explored the potential impact of distributional shifts in observational data compared to training data for the inference of dark matter mass function parameters from strong gravitational lensing data using two different machine learning frameworks (NRE and SNPE). We found that in both cases, plausible shifts in the test data (e.g. the morphology of the background sources) can negatively impact the posteriors, introducing non-negligible biases. 

When faced with out-of-distribution data, the performance of NREs and SNPEs degrades in unpredictable ways, making it difficult to calibrate the posteriors.
These tests induce a bias in the predictions of neural networks, similar to those caused by adversarial attacks \citep[e.g.,][]{Adversarial_attacks_2013}. However, note that unlike the realization of adversarial attacks, which can be statistically unlikely in real observational settings, these tests are physically motivated and are meant to represent realistic scenarios of distributional shifts. 

Nevertheless, note that even despite these challenges, NREs and SNPEs remain powerful tools in situations where distributional shifts are not a concern, including forecasting and sensitivity analysis based on in-distribution simulations. The challenge of addressing distributional shifts is already a major topic in machine learning \citep[e.g.,][]{ Masserano_2022_waldo, Montel_2022, Falkiewicz_2023_sbi_misspec, Missa_2024_Desi_dataset, Masserano_2024_likelihoodfree, Flovik_2024, Wehenkel_2024_misspec_sbi, Dellaporta_2024_robust_optim} and extensive research in domain adaptation and generalization is being pursued to alleviate these problems. For example, for certain strong lensing analysis frameworks, \citet{DomainAdaption_01} and \citet{DomainAdaption_02} have proposed domain adaptation methods to reduce these biases caused by distributional shifts in observational noise.  

These tests are also an important lesson beyond the field of strong gravitational lensing, indicating that for inference problems where high accuracy is needed but the true data-generating process is not perfectly known, NRE and SNPE results should be examined with caution.
In conclusion, while these frameworks hold great promise for specific applications, their use with real observational data demands careful consideration of inherent limitations to ensure reliable and unbiased scientific discoveries.

\begin{acknowledgments}
\section*{Acknowledgments}
This work is partially supported by Schmidt Sciences, a philanthropic initiative founded by Eric and Wendy Schmidt as part of the Virtual Institute for Astrophysics (VIA). The work is in part supported by computational resources provided by Calcul Quebec and the Digital Research Alliance of Canada. Y.H. and L.P. acknowledge support from the Canada Research Chairs Program, the National Sciences and Engineering Council of Canada through grants RGPIN-2020-05073 and 05102, and the Fonds de recherche du Québec through grants 2022-NC-301305 and 300397.
A.F. acknowledges the support from the Bourse J. Armand Bombardier.
\end{acknowledgments}

\newpage

\bibliography{Bibliography}{}
\bibliographystyle{aasjournal}

\appendix
\section{Neural Likelihood-Free Inference Methods}
\label{app:nlfim}

\subsection{Overview of Simulation-Based Inference}
In many scientific applications, it is challenging to obtain a tractable likelihood function $p(x|\theta)$ for observed data $x$ given parameters $\theta$. Instead, one often has access to a high-fidelity simulator capable of generating synthetic data $x_{\mathrm{sim}} \sim p_{\mathrm{sim}}(x |\theta)$. This problem has motivated a class of techniques collectively known as \textit{simulation-based inference} or \textit{likelihood-free inference} \citep[e.g.,][]{Papamakarios_2016, Papamakarios_2018, LueckmannEtAl2018, CranmerEtAl2019}, which aim to perform Bayesian inference directly from simulations without the need for a traceable likelihood. Historically, one of the most common strategies has been \textit{Approximate Bayesian Computation} (ABC) \citep[e.g.,][]{Beaumont_ABC_2008, Rezende_ABC_2014, Bonassi_ABC_2015}, but in high-dimensional settings ABC can become prohibitively expensive, spurring the development of more scalable likelihood-free inference approaches. In recent years, machine learning models, such as neural likelihood estimators, neural ratio estimators, and neural posterior estimators have emerged as powerful tools to achieve the likelihood-free inference goals efficiently.

\subsection{Neural Ratio Estimators (NRE)}
\label{app:nre}
Neural ratio estimators (NREs) \citep[e.g.,][]{Cranmer_2015_NRE, Brehmer_NRE_particle_2020, Brehmer_Sidd_2019_NRE, Sidd_2022_NRE, Eve_NRE_2023} are a subclass of simulation-based inference methods. 
Rather than trying to approximate the likelihood $p(x | \theta)$ or the posterior $p(\theta | x)$ directly, NREs focus on learning the likelihood-to-evidence ratio,
\begin{equation}
    r_{\phi}(x,\theta) = \frac{p_{\phi}(x | \theta)}{p(x)}.
\end{equation}
One can train a neural network to discriminate between pairs $(x,\theta)$ drawn from the joint distribution $p(\theta)\,p_{\mathrm{sim}}(x | \theta)$ and those drawn from some reference distribution. 
By doing so, the network learns to approximate $r_{\phi}(x,\theta) = p(x | \theta) / p(x)$. 
This ratio becomes useful when we consider Bayes’ rule:
\begin{equation} \label{eq:bayes_rule}
    p(\theta \mid x) \;=\; \frac{p(x \mid \theta) \, p(\theta)}{p(x)}.
\end{equation}
With the learned ratio and the prior $p(\theta)$ on the parameters, one can recover the approximate posterior:
\begin{equation}
    p(\theta | x) \;\propto\; r_{\phi}(x,\theta)\,p(\theta).
\end{equation}

\subsection{Sequential Neural Posterior Estimators (SNPE)}
\label{app:snpe}
Sequential neural posterior estimation (SNPE) \citep[e.g.,][]{He_2016_NPE, Wiqvist_2021, Wagner_C2023, Wagner_C2024} extends the idea of simulation-based inference to an iterative procedure.
The first parameters to create the simulations are drawn from a proposal distribution (in the first round the prior). These simulations are used to train a neural network to approximate the posterior directly, and using the trained network the proposal distribution is being updated to focus on the high-likelihood regions. These steps are repeated until convergence. 
By adaptively narrowing in on the region of parameter space with significant posterior mass, SNPE can be more sample-efficient than a single-shot approach.

\section{Bias Quantification}
\label{app:TARP}
In the main text, we show that the posteriors on the substructure mass function parameters obtained with an NRE - for test data out of distribution with the training data - are biased in a way that becomes evident when combining the signal of multiple lensed images. 

\begin{figure}
    \centering
    \includegraphics[width=0.24\textwidth]{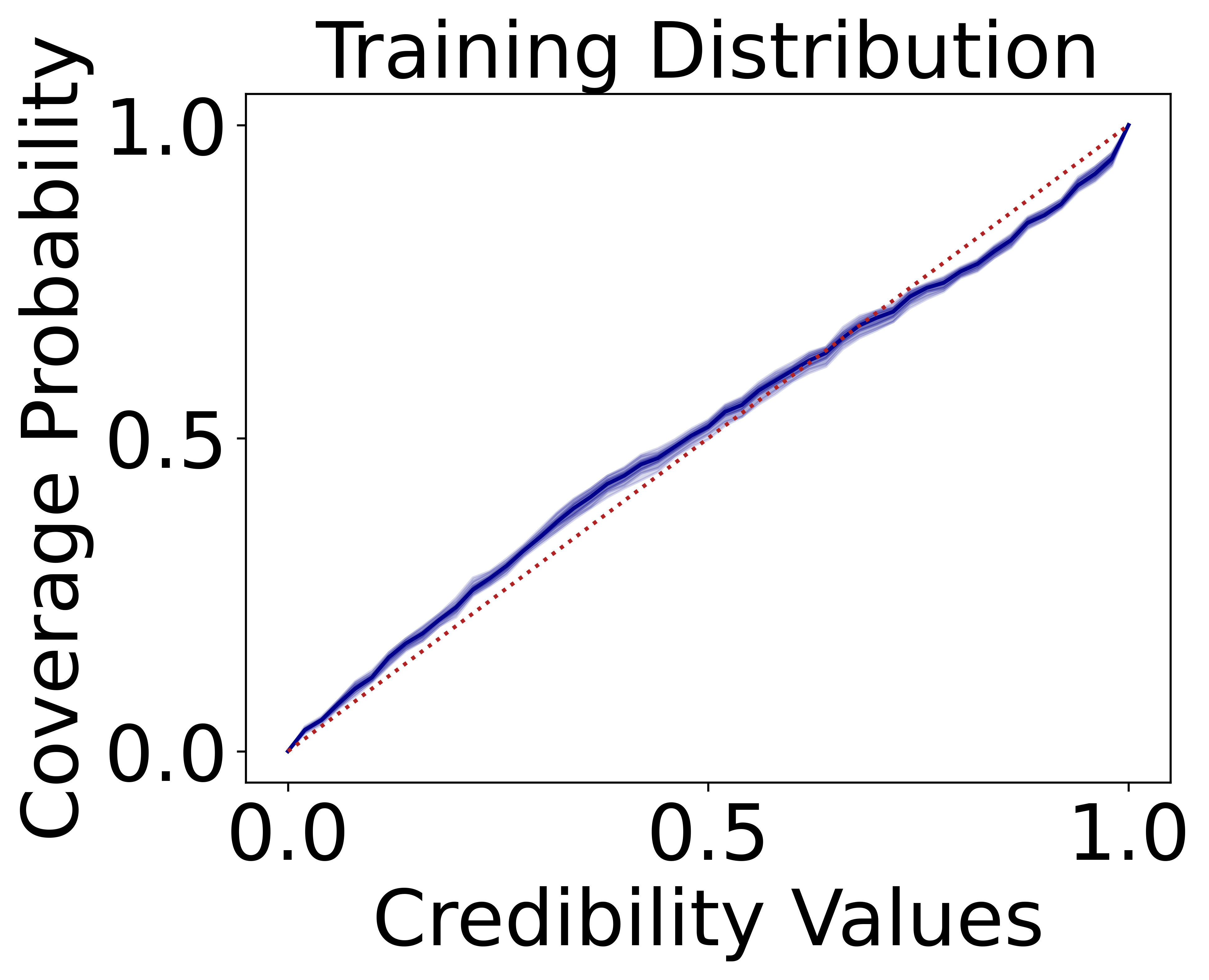}
    \includegraphics[width=0.24\textwidth]{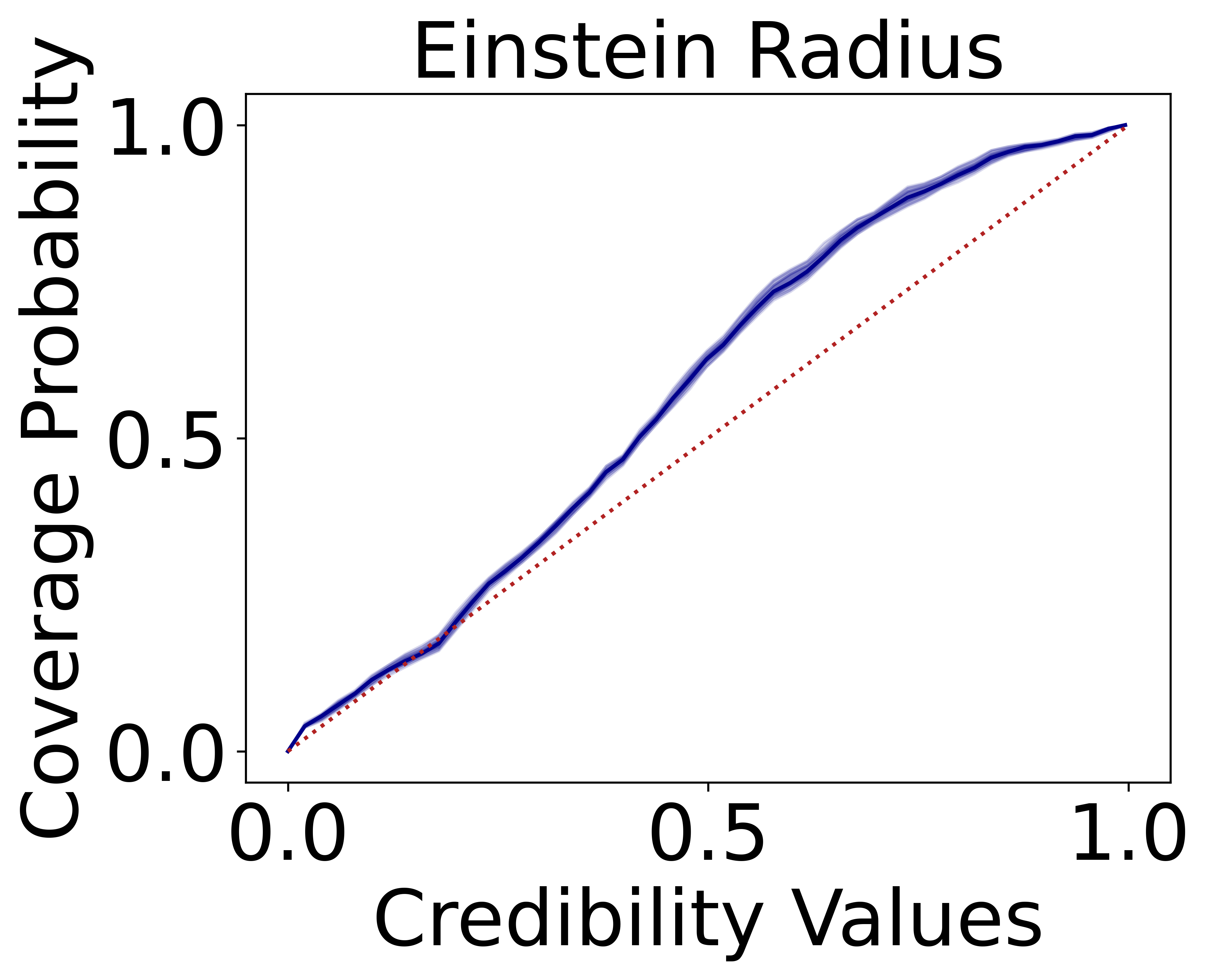}
    \includegraphics[width=0.24\textwidth]{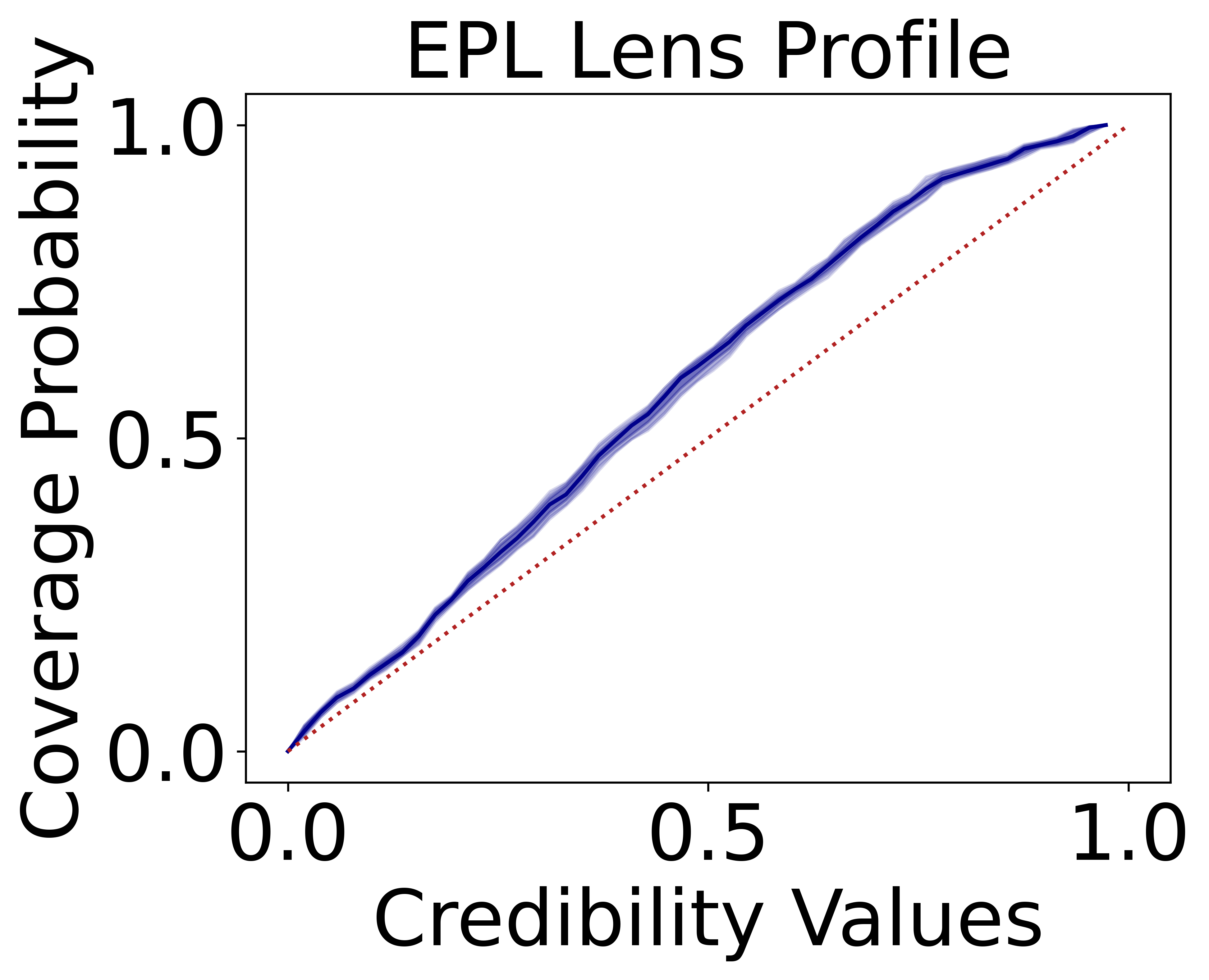}
    \includegraphics[width=0.24\textwidth]{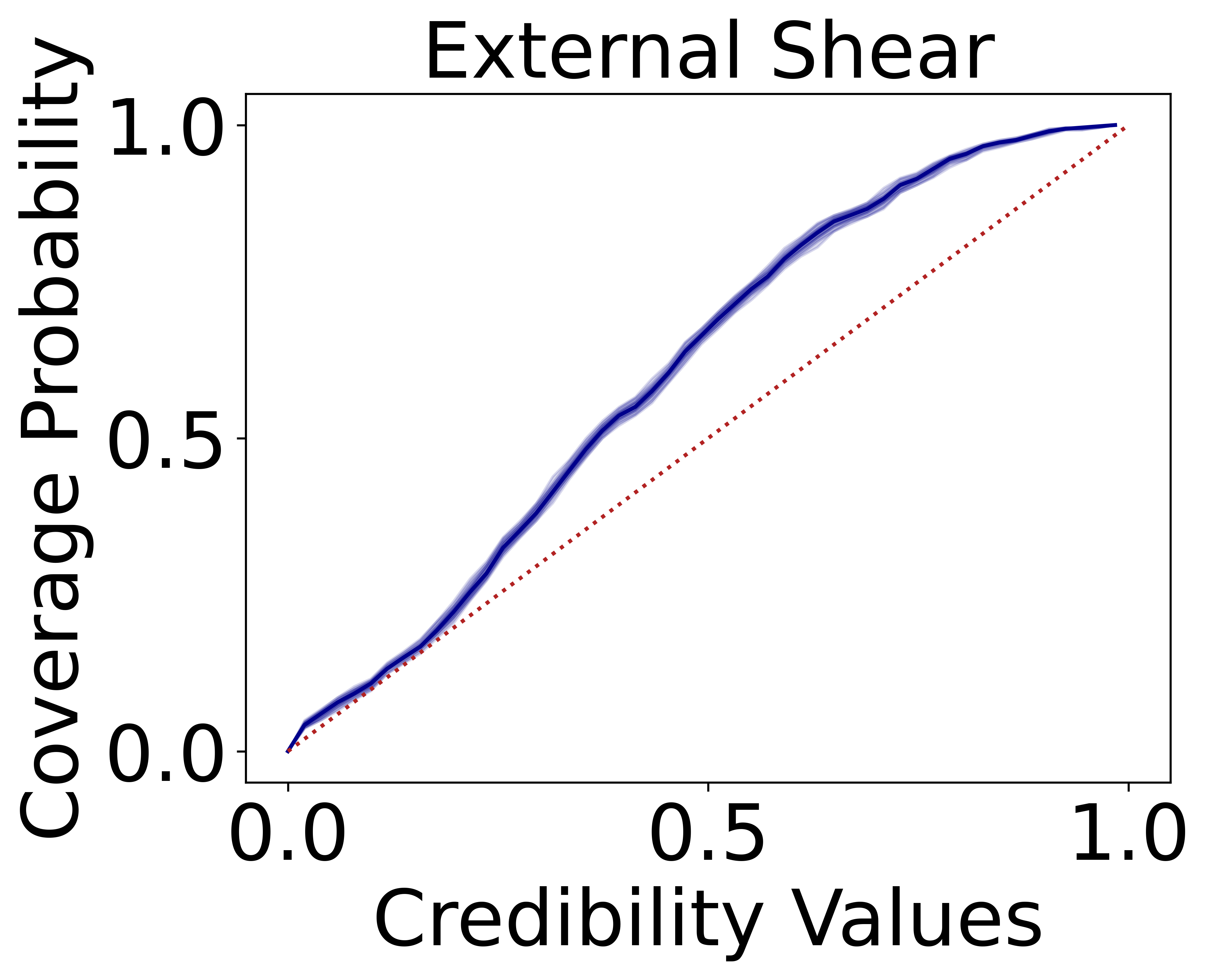}
    \includegraphics[width=0.24\textwidth]{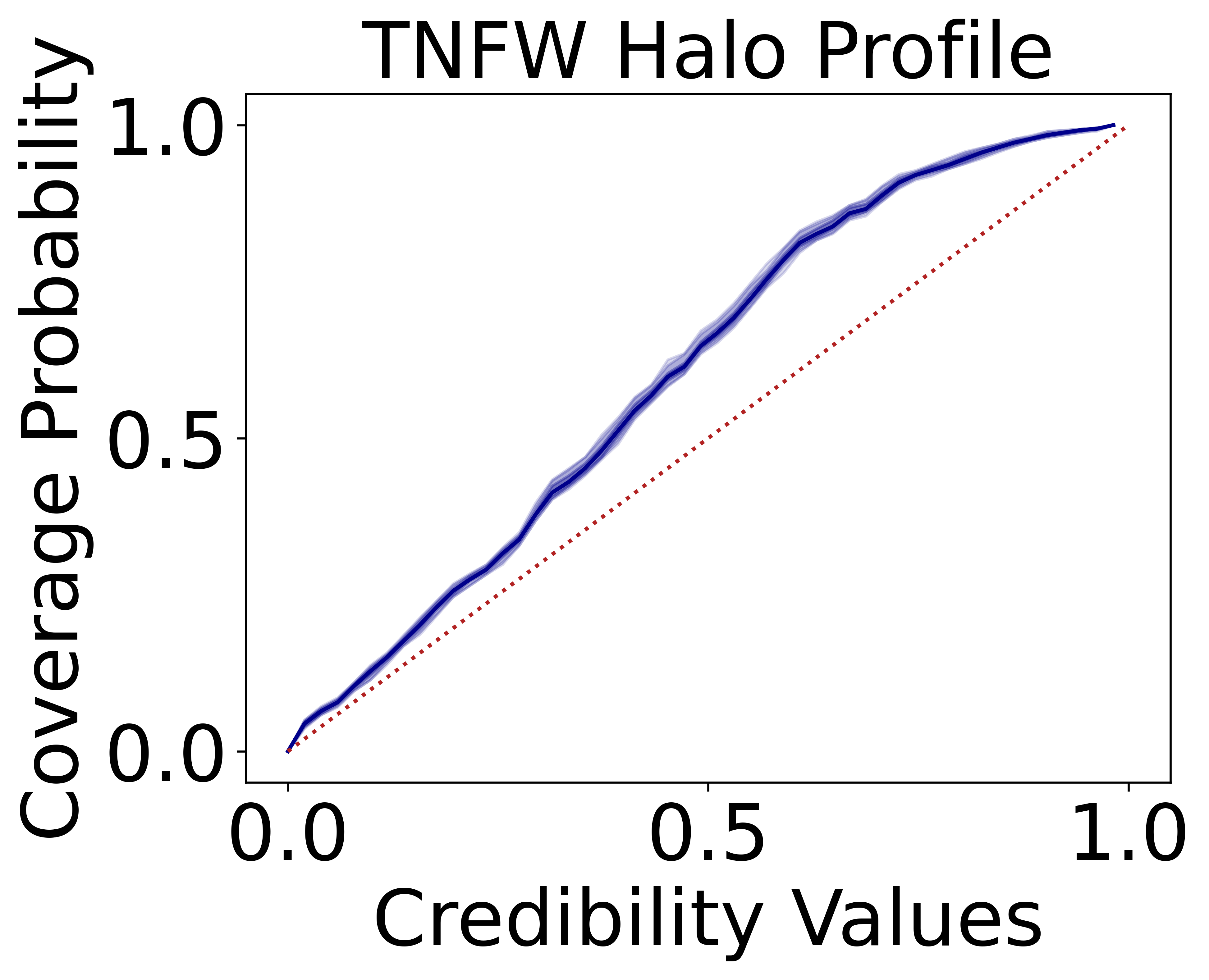}
    \includegraphics[width=0.24\textwidth]{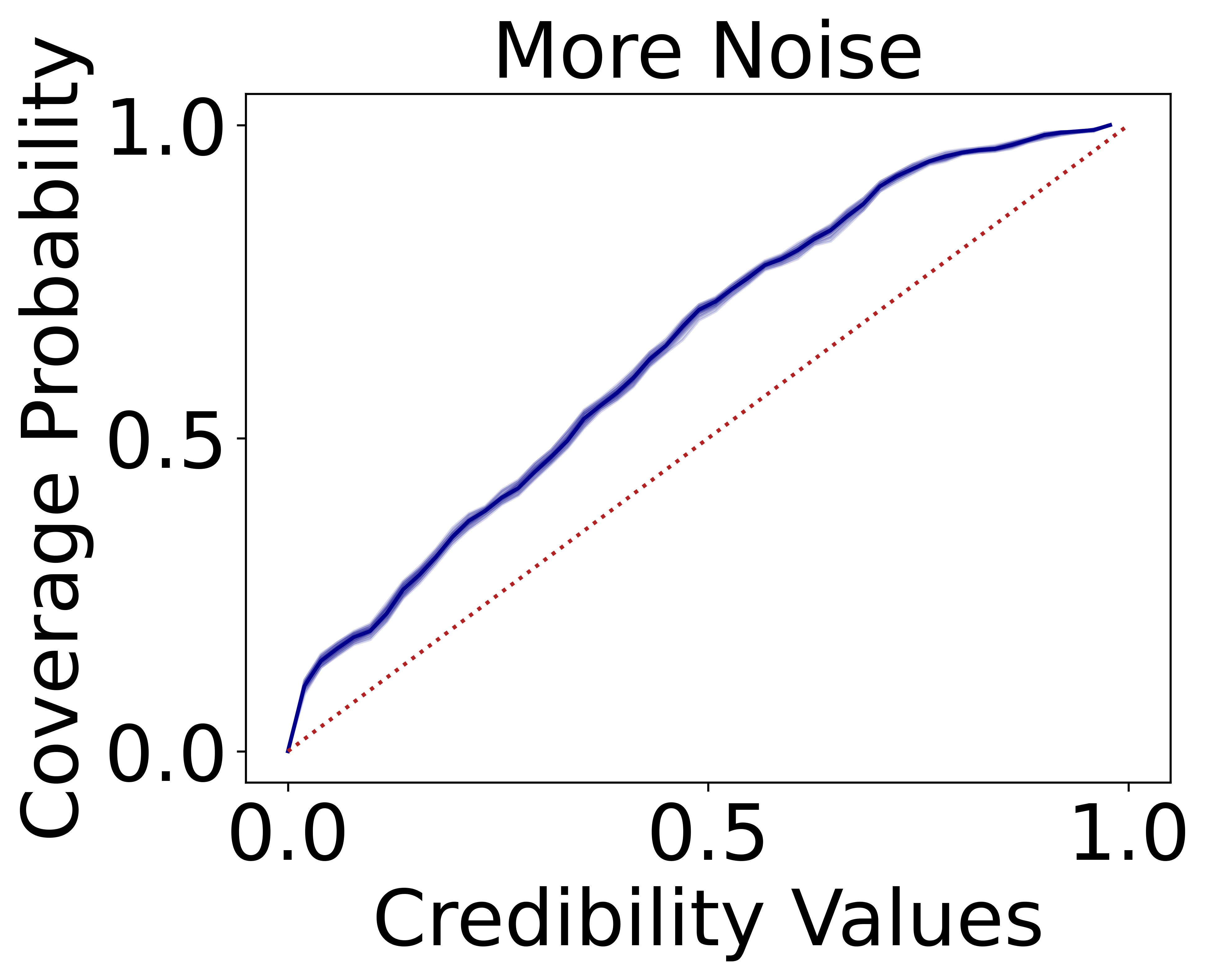}
    \includegraphics[width=0.24\textwidth]{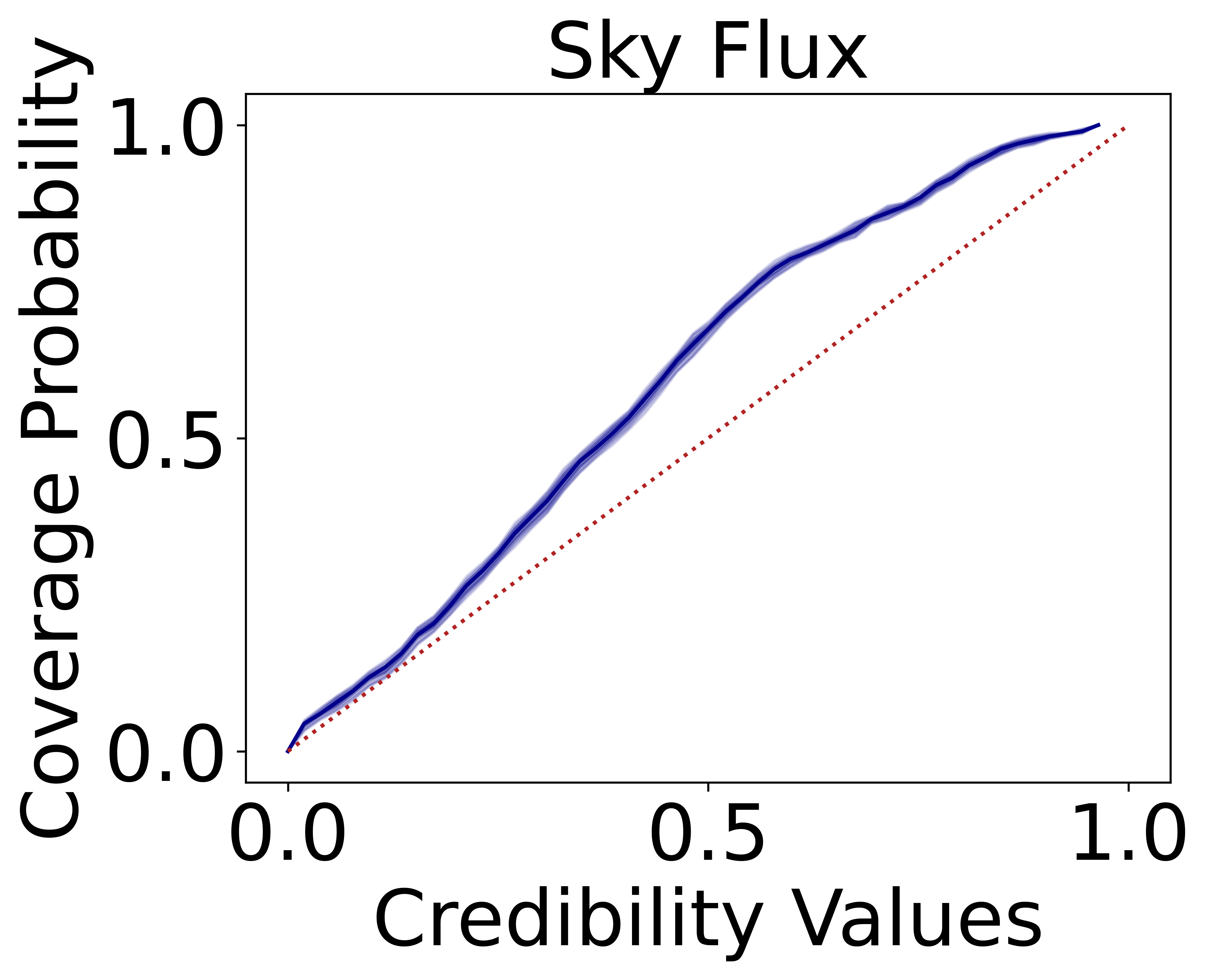}
    \includegraphics[width=0.24\textwidth]{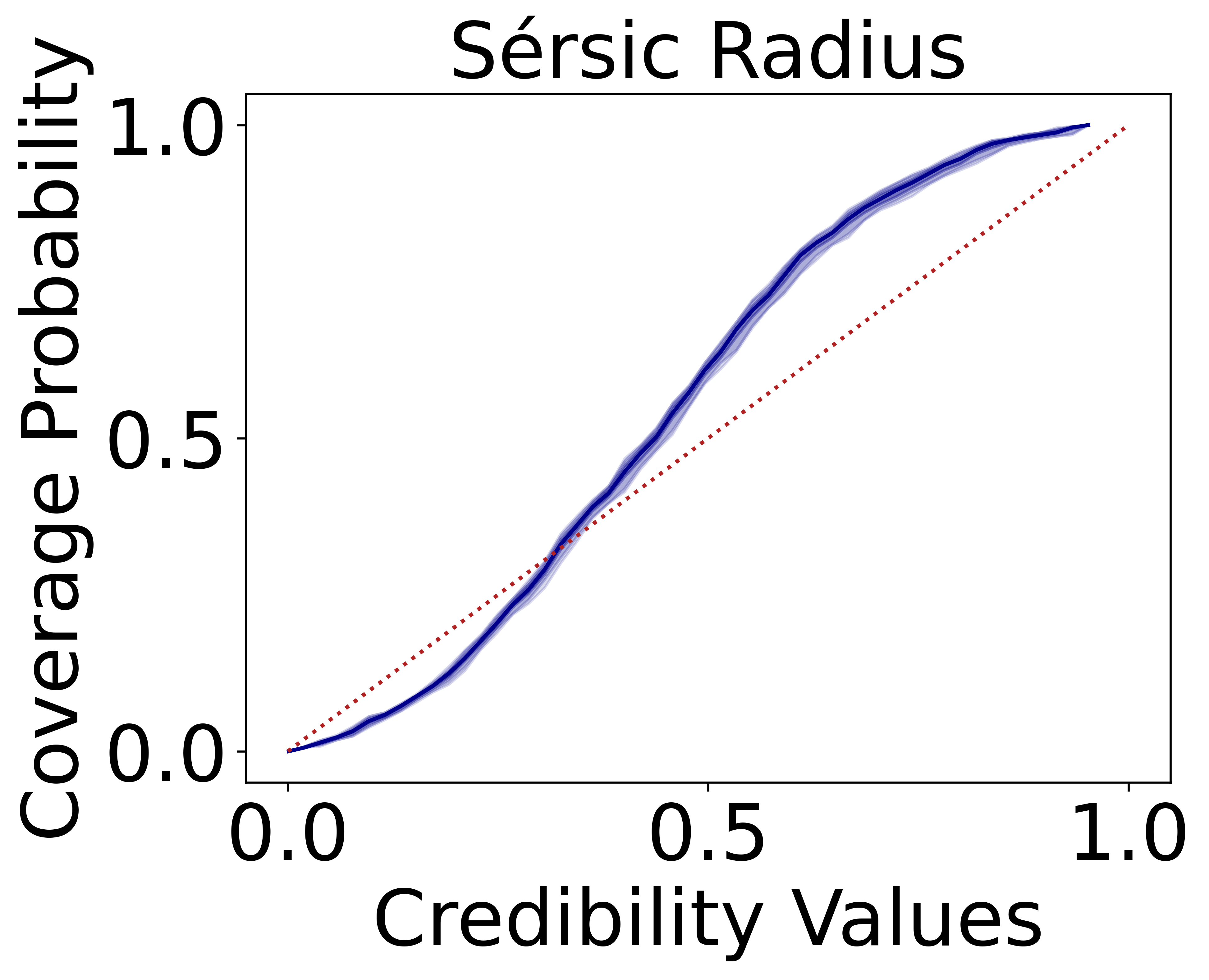}
    \includegraphics[width=0.24\textwidth]{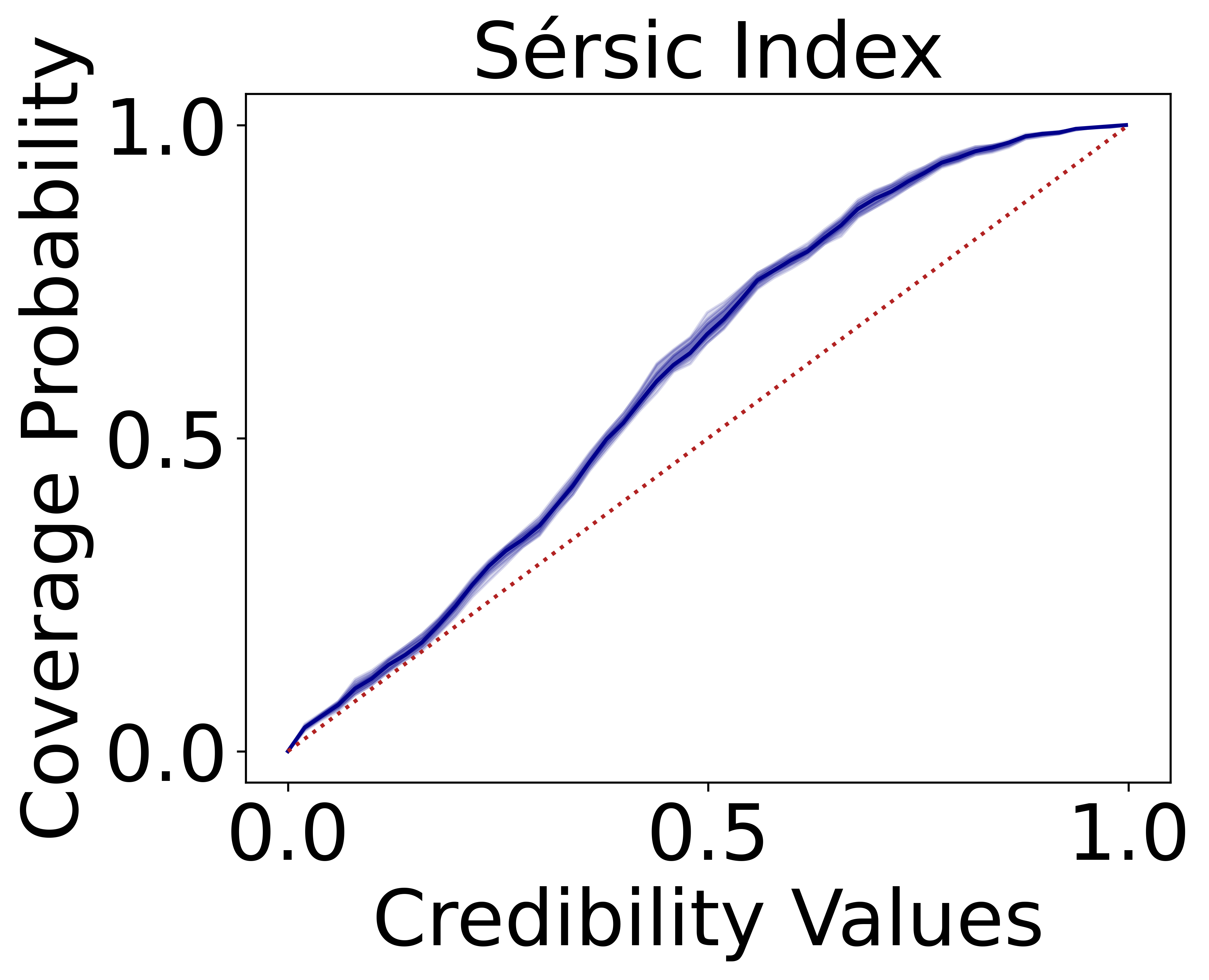}
    \includegraphics[width=0.24\textwidth]{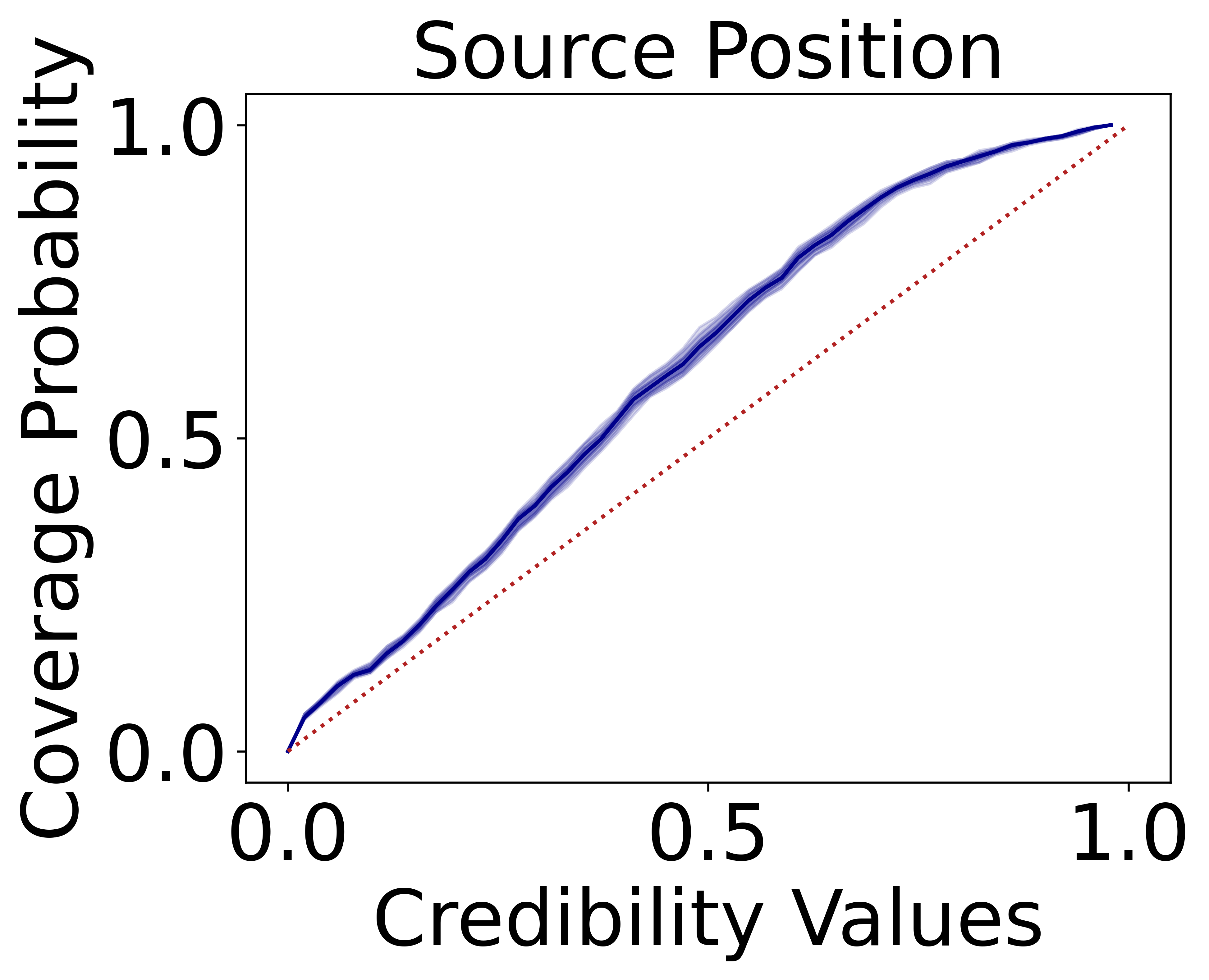}
    \includegraphics[width=0.24\textwidth]{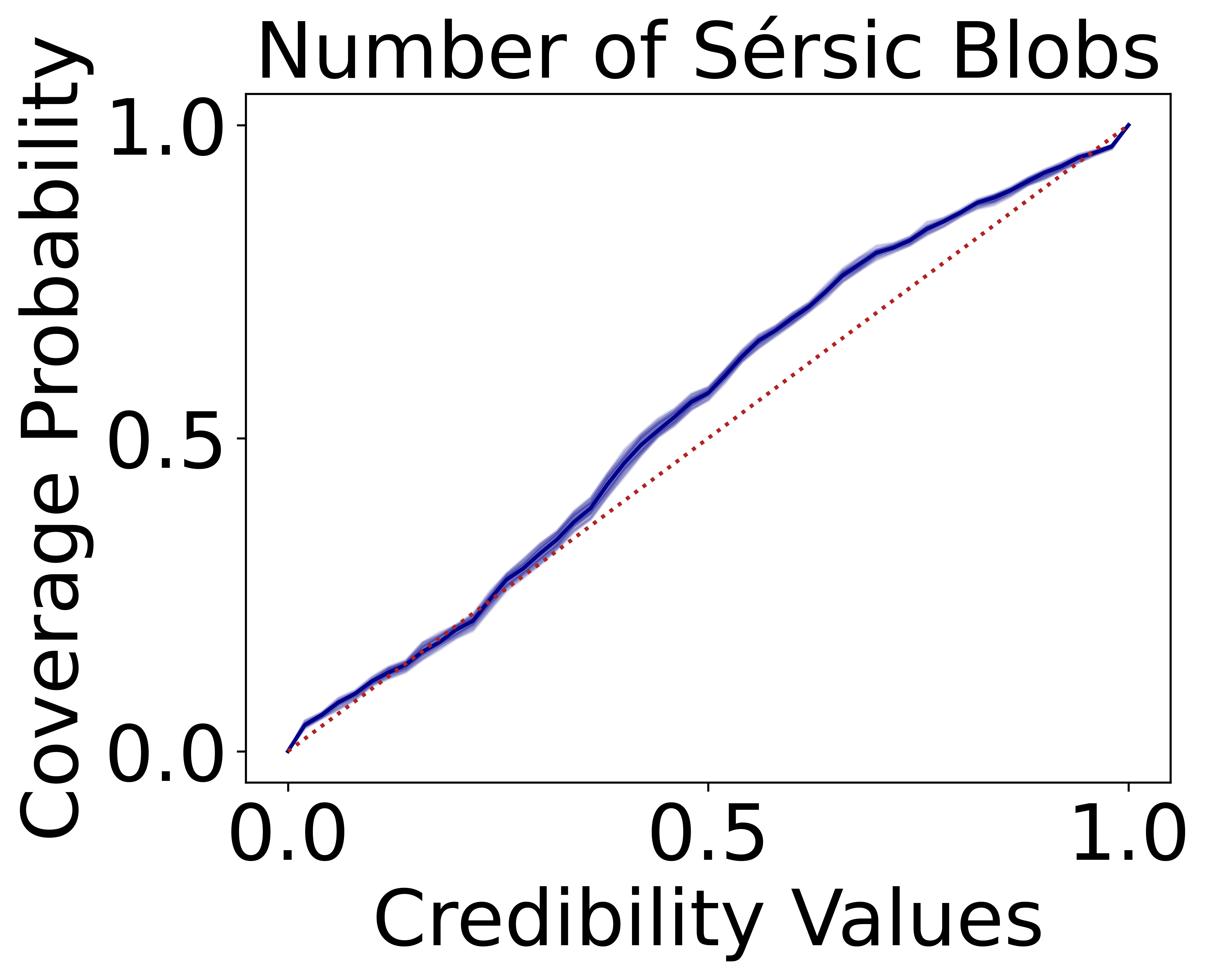}
    \includegraphics[width=0.24\textwidth]{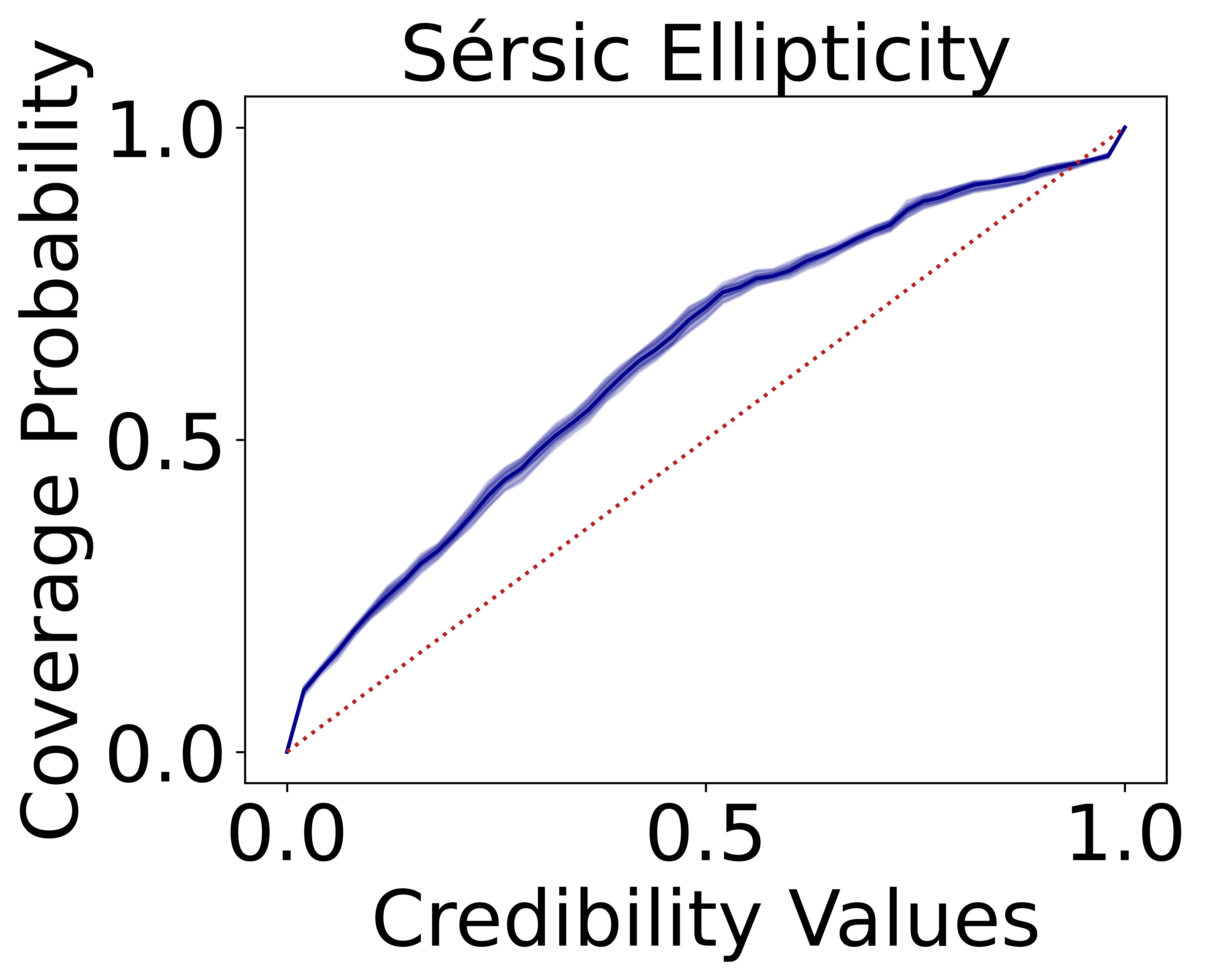}
    \includegraphics[width=0.24\textwidth]{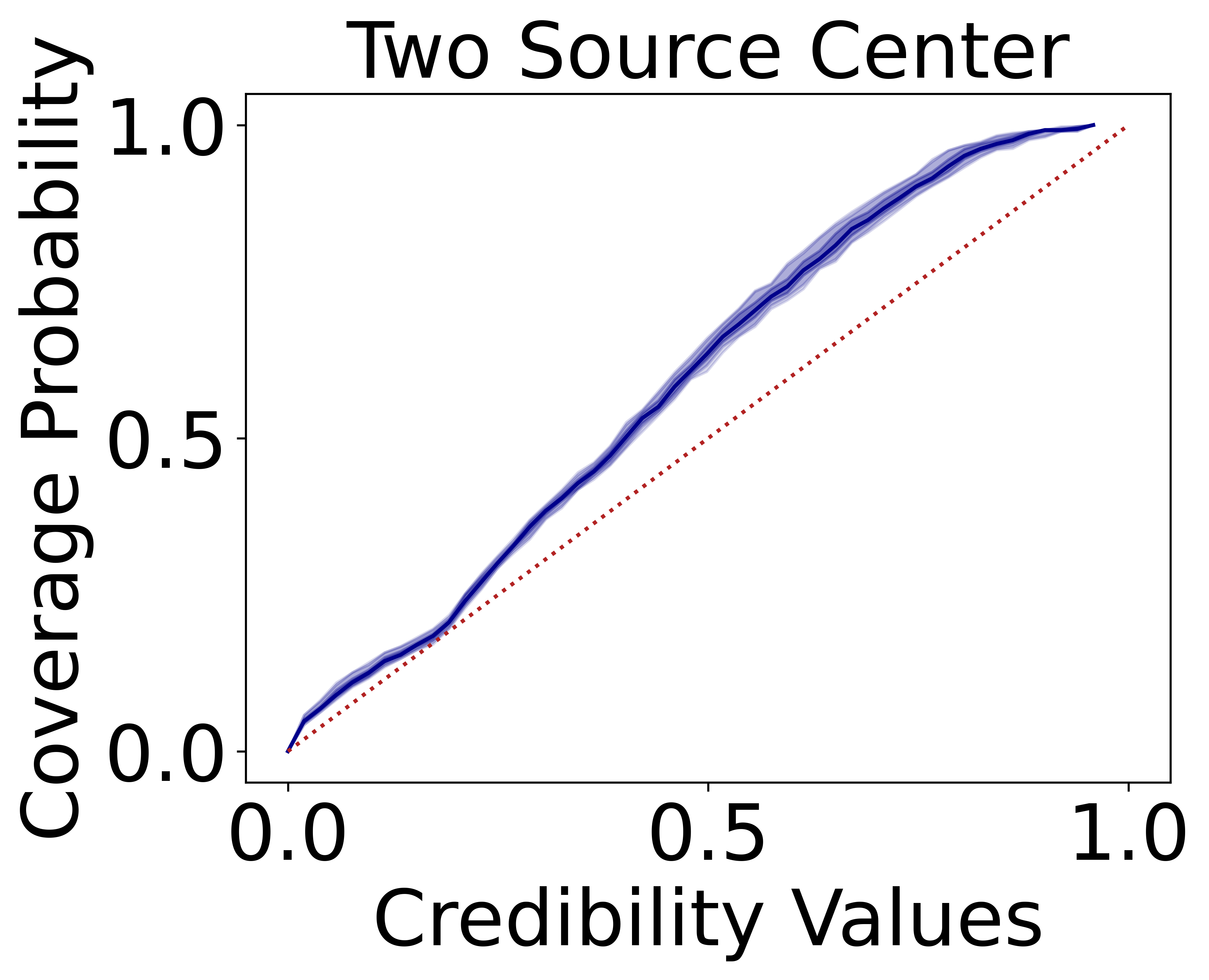}
    \includegraphics[width=0.24\textwidth]{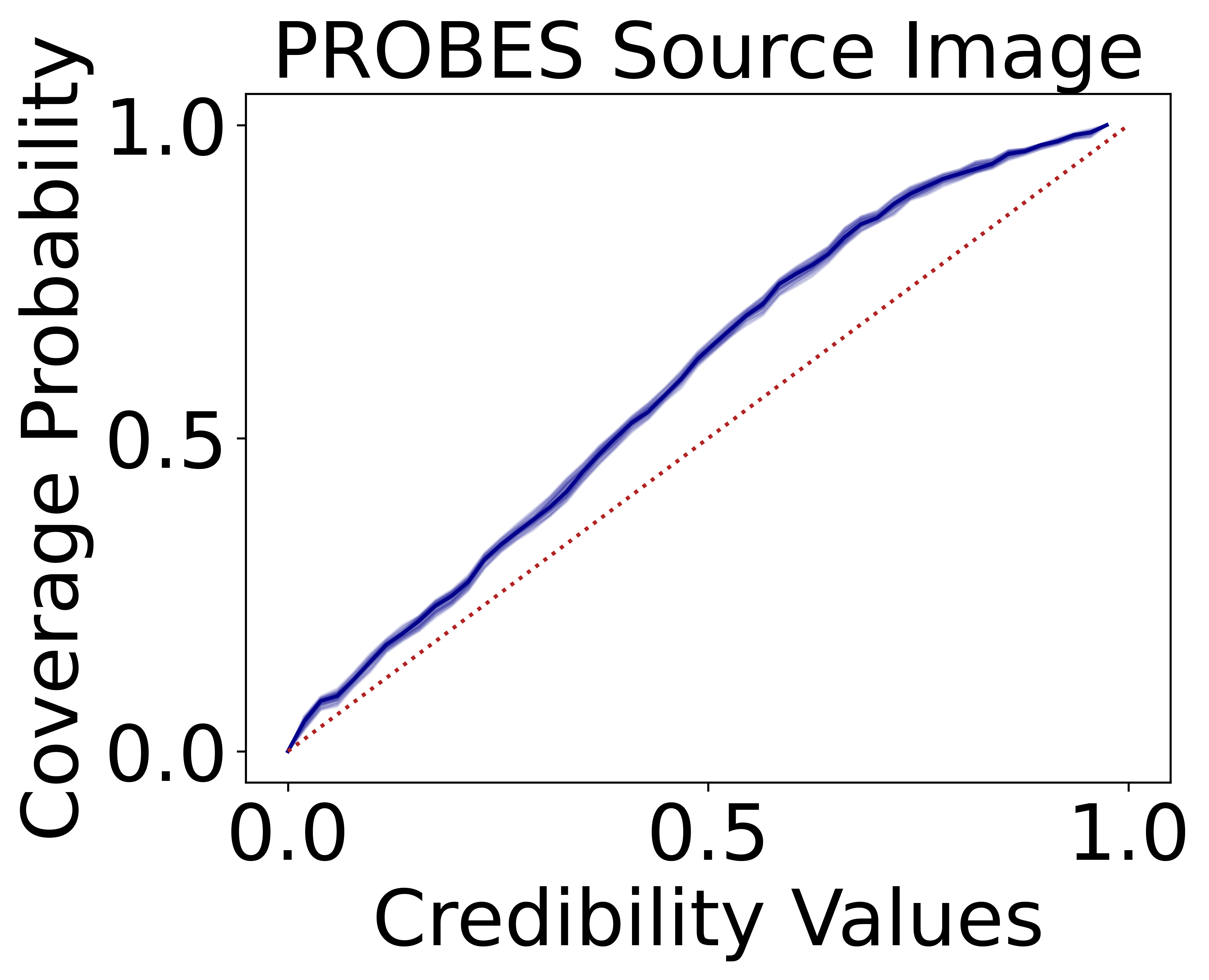}
    \caption{TARP results to quantify the per lens bias under the different tests of the NRE. The only unbiased inference results on individual-lens basis are obtained when the test distribution aligns with the training distribution.}
    \label{fig:TARP}
\end{figure}

Here, in Figure~\ref{fig:TARP} we show that this bias is measurable even on an individual-lens basis, and quantify this bias using TARP~\citep[][]{Lemos_2023_TARP}. TARP assesses the individual, on a single system level, bias on the parameters of interest rather than on the population level. These results show that the only unbiased inference results are obtained when the test distribution follows the training distribution. TARP does not provide a qualitative assessment of the biases; the results only show whether the posteriors are biased or not.
The titles of the individual TARP results follow those of Figure~\ref{fig:variations}.

\end{document}